\documentclass[traditabstract]{aa}
\usepackage{lmodern}
\usepackage{graphicx}
\usepackage{subfig}
\usepackage{natbib}
\usepackage{subfig}
\usepackage{tabularx}
\bibpunct{(}{)}{;}{a}{}{,}

\usepackage[colorlinks=true, citecolor=blue, linkcolor=blue, breaklinks=true]{hyperref}

\usepackage[fleqn]{amsmath} % \dfrac, \gamma, ...
\usepackage{amsfonts} % \mathbb \forall
\usepackage{amstext}
\usepackage{amssymb} % ulteriori simboli (ex: \blacksquare)
\newcommand{\hi } {{\rm H}\,{\small\rm I} \,}
\newcommand{\hiA} {{\rm H}\,{\small\rm I}}

\newcommand{\hiiA} {{\rm H}\,{\small\rm II}}
\newcommand{\NII } {[{\rm N}\,{\small\rm II}] \,}
\newcommand{\NIIA} {[{\rm N}\,{\small\rm II}]}
\newcommand{\avghi } {\overline{\Sigma}_{\hi}}

\begin{document}

\title{Evolution of dwarf galaxies: a dynamical perspective}
\author{Federico Lelli\inst{1}
\and Filippo Fraternali\inst{2,}\inst{1}
\and Marc Verheijen\inst{1}}

\institute{Kapteyn Astronomical Institute, University of Groningen, Postbus 800, 9700 AV, Groningen, The Netherlands \\
\email{lelli@astro.rug.nl}
\and Department of Physics and Astronomy, University of Bologna, via Berti Pichat 6/2, 40127, Bologna, Italy}

\date{}

\abstract{
For a rotating galaxy, the inner circular-velocity gradient $d_{R}V(0)$
provides a direct estimate of the central dynamical mass density,
including gas, stars, and dark matter. We consider 60 low-mass
galaxies with high-quality \hi and/or stellar rotation curves
(including starbursting dwarfs, irregulars, and spheroidals), and
estimate $d_{R}V(0)$ as $V_{R_{\rm d}}/R_{\rm d}$, where $R_{\rm d}$
is the galaxy scale-length. For gas-rich dwarfs, we find that
$V_{R_{\rm d}}/R_{\rm d}$ correlates with the central surface
brightness $\mu_{0}$, the mean atomic gas surface density
$\Sigma_{\rm gas}$, and the star formation rate surface density
$\Sigma_{\rm SFR}$. Starbursting galaxies, such as blue compact
dwarfs (BCDs), generally have higher values of $V_{R_{\rm d}}
/R_{\rm d}$ than dwarf irregulars, suggesting that the starburst
is closely related to the inner shape of the potential well.
There are, however, some ``compact'' irregulars with values of
$V_{R_{\rm d}}/R_{\rm d}$ similar to BCDs. Unless a redistribution
of mass takes place, BCDs must evolve into compact irregulars.
Rotating spheroidals in the Virgo cluster follow the same correlation
between $V_{R_{\rm d}}/R_{\rm d}$ and $\mu_{0}$ as gas-rich dwarfs.
They have values of $V_{R_{\rm d}}/R_{\rm d}$ comparable to those
of BCDs and compact irregulars, pointing at evolutionary links
between these types of dwarfs. Finally, we find that, similarly
to spiral galaxies and massive starbursts, the star-formation
activity in dwarfs can be parametrized as $\Sigma_{\rm SFR} = \epsilon
\, \Sigma_{\rm gas}/\tau_{\rm orb}$, where $\tau_{\rm orb}$ is
the orbital time and $\epsilon\simeq0.02$.
}

\titlerunning{Evolution of dwarf galaxies: a dynamical perspective}
\authorrunning{Lelli et al.}

\maketitle

\section{Introduction}\label{sec:intro}

Low-luminosity, dwarf galaxies are the most common type of galaxies in
the Universe \citep[e.g.][]{Ferguson1994}. Despite numerous observational
and theoretical studies, their formation and evolution is still not
fully understood \citep[e.g.][]{Tolstoy2009, Mayer2011, Kormendy2012}.
Three main types of dwarfs exist in the nearby Universe:
i) gas-poor dwarfs that are \textit{not} currently forming stars,
which are usually called spheroidals (Sphs) or dwarf ellipticals (dEs),
hereafter we will refer to them as Sphs; ii) gas-rich dwarfs that are
forming stars at a relatively-low rate, named irregulars (Irrs); and
iii) starbursting dwarfs that are forming stars at an unusually high
rate. The latter objects are often classified as amorphous dwarfs
(based on optical morphology, e.g. \citealt{Gallagher1987, Marlowe1999}),
\hiiA-galaxies (based on emission-line spectroscopy, e.g. \citealt
{Terlevich1991}), and/or blue compact dwarfs (BCDs, based on colors
and surface brightness measurements, e.g. \citealt{GilDePaz2003}).
Hereafter, we will refer to any starbursting dwarf as a BCD.
As we will show in Sect.~\ref{sec:SFdwarf}, the term ``BCD''
captures a fundamental observational fact: the starburst activity
(the \textit{blue} color) occurs mainly in galaxies with a steep
gravitational potential (i.e., a \textit{compact} mass
distribution towards the galaxy center), providing that they
have also a strong concentration of gas.

It is known that Sphs, Irrs, and BCDs follow the same correlations
between the effective surface brightness $\mu_{\rm eff}$, the
effective radius $R_{\rm eff}$, and the total magnitude $M$,
pointing at evolutionary links between them \citep[e.g.][]
{Kormendy1985, Binggeli1994, Tolstoy2009}. In this respect, BCDs are
particularly interesting as the burst durations are typically of the
order of a few 100~Myr \citep{McQuinn2010a}, thus they must evolve
into another type of dwarf as the starburst fades. The possibility
of morphological transformations between low-mass galaxies is also
suggested by the existence of ``transition type'' dwarfs, which have
intermediate properties between Sphs and Irrs/BCDs \citep[e.g.][]
{Sandage1991, Mateo1998, Dellenbusch2007, Dellenbusch2008}.

Several photometric studies have shown that the underlying, old stellar
component of BCDs typically has a smaller scale-length and a higher
central surface brightness than Irrs and Sphs of the same luminosity,
suggesting that the evolutionary links between BCDs and Irrs/Sphs are
not straightforward \citep[e.g.][]{Papaderos1996, GilDePaz2005, Herrmann2013}.
However, it is generally difficult to obtain accurate structural parameters
for starbursting dwarfs, as the galaxy morphology is extremely irregular
and young stars may dominate the integrated light over much of the
stellar body. Recently, \citet{Micheva2013} obtained deep optical
and near-infrared photometry, and challenged the previous results,
arguing that the structural parameters of the old stellar component
of BCDs are consistent with those of Irrs and Sphs.

A different approach is to consider dynamical information that directly
traces the distribution of mass, such as \hi rotation curves \citep[e.g.]
[]{Lelli2012a, Lelli2012b}. Using qualitative estimates of the rotation
velocities, \citet{vanZee2001} suggested that BCDs have steeper rotation
curves than low surface brightness galaxies of similar luminosity
\citep[see also][]{Meurer1998}. In \citet{Lelli2012b}, we considered
a small sample of BCDs and Irrs with high-quality \hi rotation curves,
and measured the circular-velocity gradient $V_{R_{\rm d}}/R_{\rm d}$,
where $R_{\rm d}$ is the exponential scale-length of the stellar body.
We found that BCDs generally have higher values of $V_{R_{\rm d}}
/R_{\rm d}$ than typical Irrs, implying that they have a higher central
dynamical mass density (including gas, stars, and dark matter). BCDs
also have higher central \hi surface densities than Irrs \citep[e.g.][]
{vanZee1998, vanZee2001, Simpson2000}. This suggests that the starburst
is closely related to the inner shape of the gravitational potential and
to the central concentration of gas. This connection must be the key to
understanding the mechanisms that trigger and drive the starburst in BCDs.

In this paper, we confirm the results of \citet{Lelli2012b} for a larger
sample of BCDs and Irrs, and include star formation rate (SFR) indicators
in the analysis. We also consider a sample of rotating Sphs. We use the
dynamical information provided by $V_{R_{\rm d}}/R_{\rm d}$ to constrain
the possible evolutionary links between dwarf galaxies.

\section{The sample}

We define a dwarf galaxy, in a dynamical sense, as an object with
$V_{\rm flat}\leq100$ km~s$^{-1}$, where $V_{\rm flat}$ is the asymptotic
velocity along the flat part of the rotation curve. For a pressure-supported
system, $V_{\rm flat}$ can be estimated as $\sqrt{3} \sigma_{\rm obs}$ \citep{McGaugh2010},
where $\sigma_{\rm obs}$ is the observed velocity dispersion along the line
of sight. According to the Tully-Fisher (TF) relation, $V_{\rm{flat}}\simeq100$
km~s$^{-1}$ occurs at $M_{\rm{B}}\simeq -16.5$ mag \citep[cf.][]{Verheijen2001b},
thus our definition of a dwarf galaxy qualitatively agrees with the standard
one given by \citet{Tammann1994}, which is based on total luminosity and
size. However, contrary to Tammann's criteria, our definition is directly
related to the potential well of the galaxy and is less affected by
the effects of recent star-formation, which can be serious for BCDs where
the light is dominated by young stellar populations. The choice of 100
km~s$^{-1}$ is \textit{not} arbitrary: in galaxies with $V_{\rm flat}
\leq 100$ km~s$^{-1}$ bulges tend to disappear \citep[e.g.][]{Kormendy2012}
and some cosmological models predict that mass loss from supernova feedback
may start to affect the baryonic content \citep[e.g.][]{Dekel1986}. Using
the baryonic TF relation  \citep[e.g.][]{McGaugh2012}, we estimate that
galaxies with $V_{\rm flat}\leq 100$ km~s$^{-1}$ have a baryonic mass
(stars and atomic gas) $M_{\rm bar}\lesssim 5 \times 10^9$~$M_\odot$.

We built a sample of dwarf galaxies with high-quality rotation curves, 
retrieving optical and \hi data from various sources. We included in
our selection also galaxies with rotation curves that do not reach the
flat part but have $V_{\rm last} < 100$ km~s$^{-1}$, where $V_{\rm last}$
is the circular velocity at the last measured point. The dynamical
masses of these objects are uncertain, as their rotation curves may
continue to rise, but their total magnitudes are $\lesssim -18$ R mag
($\lesssim -17$ B mag), indicating that these galaxies are actual
dwarfs. In the following, we describe our sub-samples of starbursting
dwarfs (BCDs), typical star-forming dwarfs (Irrs), and gas-poor dwarfs
(Sphs). We also clarify the nomenclature used throughout this paper.

\subsection{Starbursting dwarfs}

In Lelli et al. (submitted), we built a sample of 18 starbursting dwarfs by
considering objects that satisfy two criteria: i) they have been resolved into
single stars by the \textit{Hubble Space Telescope} (HST); and ii) their star
formation histories (SFHs), as derived by modelling color-magnitude diagrams
(e.g. \citealt{McQuinn2010a}), show an increase in the recent SFR
($\lesssim$1~Gyr) by a factor $\gtrsim$~3 with respect to the average,
past SFR. We consider here a sub-sample of 8 objects, for which
\hi rotation curves could be derived (see Lelli et al., submitted). For another
object (SBS~1415+437), we could derive a rotation curve but this is quite
uncertain and may not be a reliable tracer of the gravitational potential,
as the galaxy strongly deviates from the baryonic TF relation (see Fig.~8 in
Lelli et al., submitted), thus we exclude this object here. As we stressed
in Sect.~\ref{sec:intro}, we refer to any starbursting dwarf as a BCD.

We also added the well-studied BCD NGC~2915, which has been resolved
into single stars by HST \citep{Karachentsev2003}, but its SFH has
not yet been derived. NGC~2915 has a regularly-rotating \hi disk
\citep{Elson2010}, but the inner parts of the rotation curve are
uncertain because of the presence of strong non-circular motions
\citep{Elson2011}, thus we assigned a conservative error of 15
km~s$^{-1}$ to the inner points of the rotation curve, which have
rotation velocities between $\sim$30 and $\sim$60 km~s$^{-1}$.

The properties of our sample of 9 BCDs are given in Tables \ref{tab:BCDs}
and \ref{tab:BCDs2}. For all these objects, the HST studies provide
accurate distances using the tip of the red giant branch (TRGB)
method. For NGC~2366, we used the distance derived by
\citet{Tolstoy1995} from Cepheids observations, which is consistent
within the uncertainties with that obtained from the TRGB.

\subsection{Irregulars}

We selected 37 Irrs from the sample of \citet{Swaters2009}. We required
that the galaxies have high-quality rotation curves (quality-flag
$q\leq 2$, see \citealt{Swaters2009} for details) and inclinations between
$30^{\circ}$ and $80^{\circ}$, thus the rotation velocities and the central
surface brightnesses can be measured with small uncertainties. The rotation
curves of these galaxies have been derived by \citet{Swaters2009} taking
into account beam-smearing effects. We also added another 6 objects that
meet our quality-criteria: UGC~6955 (DDO~105) and UGC~8320 (DDO~168) from
\citet{Broeils1992}, UGC~6399 and UGC~6446 from \citet{Verheijen2001},
and the Local Group dwarfs WLM \citep{Jackson2004} and NGC~6822
\citep{Weldrake2003}. These 43 galaxies are classified as Irr, Im, Sm,
or Sd \citep{RC3}; for simplicity we refer to all of them as Irr.

It is possible that some of these Irrs may harbour a starburst and, thus,
should be considered as BCDs. For example, the sample of \citet{Swaters2009}
contains NGC~4214 (IBm), NGC~2366 (IBm), and NGC~4068 (Im), which are part
of our sample of starbursting dwarfs. Moreover, \citet{McQuinn2010a} studied
the SFH of NGC~6822, a prototype Irr in the Local Group, and found that it may
have experienced a recent starburst. However, since the HST field-of-view covers
only $\sim$10$\%$ of the stellar body of NGC~6822 \citep{McQuinn2012}, the
SFH is representative of a small fraction of the galaxy, thus we prefer to
consider NGC~6822 as a typical Irr \citep[see also][]{Mateo1998, Tolstoy2009}.
The sample of \citet{Swaters2009} also contains Holmberg~II (UGC~4305),
which is a well-studied starbursting dwarf \citep[][]{Weisz2008, McQuinn2010b}.
For this galaxy, the value of the inclination $i$ is uncertain:
\citet{Swaters2009} assumed $i = 40^{\circ}$, \citet{Oh2011} derived $i =
49^{\circ}$ from a tilted-ring fit to the velocity field and $i = 25^{\circ}$
from the baryonic TF relation, whereas \citet{Gentile2012} constrained
the outer value of $i$ between 20$^{\circ}$ and 35$^{\circ}$ by building
3-dimensional disk models. Given these uncertainties, we chose to exclude
Holmberg~II.

Tables \ref{tab:Irrs} and \ref{tab:Irrs2} provide the properties of our sample
of 43 Irrs. Galaxy distances have been taken from the literature adopting the
following distance indicators (in order of decreasing priority): Cepheids,
TRGB, and TF relation.

\subsection{Rotating spheroidals}

The sample of gas-poor dwarfs is drawn from \citet{vanZee2004a, vanZee2004b},
who used long-slit optical spectroscopy to derive stellar rotation curves and
velocity dispersion profiles of 16 low-luminosity galaxies in the Virgo cluster.
We selected only 8 objects that show a clear rotation-velocity gradient for
$R\lesssim R_{\rm d}$. These galaxies have been classified as dEs by
\citet{Binggeli1985}. Following \citet{Kormendy2009}, we refer to them as Sphs,
since they are intrinsically different from bright elliptical galaxies and
follow different scaling relations. We assume that the Virgo cluster is at
a distance of 16.1$\pm$1.2 Mpc \citep{Kelson2000}.

Van Zee et al. (2004b) provide the rotation velocities projected along the line
of sight. To trace the gravitational potential of the galaxy, these velocities
must be corrected for inclination and pressure-support. To correct for inclination,
we modelled the Sphs as thick disks (see Sect.~\ref{sec:struct}).
To correct for pressure-support, we calculated the asymmetric-drift correction
(see Appendix \ref{sec:asymDrift}). These 8 objects have $V_{\rm{last}} \leq 100$
km~s$^{-1}$ after applying the inclination and asymmetric-drift corrections,
thus they fulfill our definition of a dwarf galaxy. The properties of our sample
of rotating Sphs are given in Table~\ref{tab:Sphs}.

\section{Data Analysis\label{sec:dataAnal}}

In Sect. \ref{sec:res}, we will present several diagrams that compare the dynamical,
structural, and star-formation properties of dwarf galaxies. Here we describe the
quantities that enter in those diagrams and provide details about the error budget.

\subsection{The circular-velocity gradient\label{sec:V/R}}

The key quantity here is the inner circular-velocity gradient 
\begin{equation}
 d_{R}V(0) = \lim_{R\rightarrow 0} \dfrac{dV_{\rm circ}(R)}{dR},
\end{equation}
where $V_{\rm circ} (R)$ is the circular velocity of a test particle
orbiting at the galactocentric radius $R$ \citep[see][]{Lelli2013}.
For a spherical distribution of mass, $d_{R}V(0) = \sqrt{4/3 \pi G\rho_{0}}$
where $G$ is Newton's constant and $\rho_{0}$ is the central dynamical
mass density (including gas, stars, and dark matter). Thus, for a given
dynamical mass, $d_{R}V(0)$ provides, to a first approximation, a direct
measure of the \textit{inner steepness} of the potential well of a galaxy.
The rotation curves of dwarf galaxies are typically described by a nearly
solid-body part for $R \lesssim 2 R_{\rm d}$ (where $R_{\rm d}$ is the
exponential scale-length) and an outer flat part for $R\gtrsim 2 R_{\rm d}$
\citep[e.g.][]{Swaters2009}. Thus, we can estimate $d_{R}V(0)$ on the
solid-body portion of the rotation curve as $V_{R_{\rm d}}/R_{\rm d}$,
where $V_{R_{\rm d}} = V_{\rm circ}(R_{\rm d})$. If high-resolution
rotation curves are available, it is possible to quantify $d_{R}V(0)$
using more sophisticated techniques, such as a polynomial fit
that takes into account the possible concave-down shape of the
inner rotation curve \citep[see][]{Lelli2013}. For bulgeless galaxies
with a nearly-exponential luminosity profile, this would result
in differences in $d_{R}V(0)$ within a factor of $\sim$2. For
the purposes of this paper, a simple estimate of $d_{R}V(0)$
as $V_{R_{\rm d}}/R_{\rm d}$ is sufficiently accurate. We also
note that, since we are considering the solid-body portion of
the rotation curve, the exact value of $R_{\rm d}$ does not strongly
affect the value of $d_{R}V(0) \simeq V_{R_{\rm d}}/R_{\rm d}$.
The rotation curves used here were derived taking into account
the possible effects of beam-smearing and non-circular motions
(see \citealt{Swaters2009, Lelli2012a, Lelli2012b} and Lelli
et al., submitted).

For gas-rich galaxies (BCDs and Irrs), $V_{\rm circ}(R)$ is directly provided
by the \hi rotation curve $v_{\rm rot}(R)$, as the asymmetric-drift correction
to account for the pressure support is typically negligible in the inner
galaxy regions \citep[e.g.][]{Swaters2009, Lelli2012b}. For gas-poor
galaxies (Sphs), instead, the stellar rotation curves must be corrected
for pressure support. The asymmetric-drift correction is described in
Appendix~\ref{sec:asymDrift}.

To estimate the error $\delta_{V/R}$ on $V_{R_{\rm d}}/R_{\rm d}$,
we consider the following equation:
\begin{equation}
\dfrac{V_{\rm{circ}}}{R} = \dfrac{v_{\rm{l.o.s}}}{\sin(i)} \dfrac{1}{\alpha D},
\end{equation}
where $v_{\rm{l.o.s}}$ is the circular-velocity projected along the line of sight,
$i$ is the inclination, $\alpha$ is the angular scale-length (in radians), and
$D$ is the galaxy distance. The propagation of the errors gives
\begin{equation}\label{eq:errGrad}
\delta_{V/R}^{2} = \bigg\{ \bigg[ \dfrac{\delta_{v_{\rm l.o.s}}}{R \sin(i)}\bigg]^{2} +
                               \bigg[ \dfrac{V_{\rm{circ}}}{R} \dfrac{\delta_{i}}{\tan(i)}\bigg]^{2} +
			       \bigg[ \dfrac{V_{\rm{circ}}}{R} \dfrac{\delta_{D}}{D}\bigg]^{2}\bigg\}_{R=R_{\rm d}}
\end{equation}
where the error on $\alpha$ has been neglected. For $\delta_{v_{\rm l.o.s}}/\sin(i)$
we used the error on the rotation velocities given in the original papers; this
includes the formal error given by a $\chi^{2}$-minimization and an additional
uncertainty due to the asymmetries between the approaching and the receding side
of the galaxy \citep[see e.g.][]{Swaters2009}. For $\delta_{i}$ we assumed a typical
error of 3$^{\circ}$, as $\delta_{i}$ is not provided in the original papers (except
in a few cases, see Tables \ref{tab:BCDs} and \ref{tab:Irrs}). Typically, $\delta_{ D}$
gives a negligible contribution for galaxies with Cepheids and TRGB distances,
whereas it dominates the error budget for galaxies with TF distances.

\subsection{Structural parameters\label{sec:struct}}

We collected $R$-band apparent magnitudes $m_{\rm R}$, central surface brightnesses
$\mu_{0}$, and scale-lengths $R_{\rm d}$ from various sources. For the BCDs, we refer
to Lelli et al. (submitted). For the Irrs, the sources are \citet{Swaters2002b} (40 galaxies),
\citet{Tully1996} (UGC~6399 and UGC~6446), and \citet{Hunter2006} (WLM and NGC~6822,
their $V$-band values have been converted to $R$-band assuming $V-R = 0.5$, cf.
with e.g. \citealt{Micheva2013}). For the Sphs, the source is \citet{vanZee2004a}.
These authors derived the structural parameters $\mu_{0}$ and $R_{\rm d}$ by fitting
an exponential function to the outer parts of the surface brightness profiles.
For BCDs, the resulting values of $\mu_{0}$ and $R_{\rm d}$ are thought to be
representative of the underlaying, old stellar component \citep[e.g.][]{Papaderos1996,
GilDePaz2005}.

We calculated absolute magnitudes $M_{\rm R}$ using our adopted distances,
and estimated the errors considering the distance uncertainties only, hence
$\delta_{M} = 5 \log(e) \delta_{D}/D$. We corrected $\mu_{0}$ for inclination
$i$ using the following equation:
\begin{equation}
\mu^{i}_{0} = \mu_{0} - 2.5 C \log[\cos(i)]
\end{equation}
where $C$ is a constant related to the internal extinction. Since dwarf galaxies
typically have low metallicities (see tables \ref{tab:BCDs2} and \ref{tab:Irrs2}),
the dust content is likely low. Hence, we assumed that they are optically-thin
($C = 1$). The error $\delta_{\mu_{0}^{i}}$ on $\mu_{0}^{i}$ is given by:
\begin{equation}
\delta_{\mu^{i}}^{2} = \delta_{\mu_{0}}^{2} + [ 2.5 \log(e) \tan(i) \delta_{i} ]^{2}.
\end{equation}
Since $\delta_{\mu_{0}}$ is usually not provided in the original papers, we assumed
that $\delta_{\mu_{0}} = 0.1$~mag. This conservative choice takes into account
uncertainties on the photometric calibration, on the exponential fits, and on the
fact that the surface brightness profiles of dwarf galaxies may have inner cores
or cusps \citep[see e.g.][]{Swaters2002b}.

For BCDs and Irrs, the inclination was derived by fitting a tilted-ring model to the
\hi velocity field and/or by building model-cubes (see Lelli et al., submitted and
\citealt{Swaters2009} for details). For Sphs, we estimated $i$ from the observed
ellipticities $\varepsilon$ using the formula
\begin{equation}\label{eq:incl}
\cos{i}^{2}= \dfrac{(1 - \varepsilon)^{2} - q_{0}^{2}}{1 - q_{0}^{2}},
\end{equation}
where $q_{0}$ is a constant that depends on the oblateness of the stellar distribution.
We assumed $q_{0} = 0.35$, as indicated by statistical studies of the observed
ellipticities of dwarf galaxies \citep{Lisker2007, Sanchez2010}. If one assumes that
$q_{0} = 0.2$, a typical value for the stellar disks of spiral galaxies \citep[e.g.]
[]{Holmberg1950}, the difference in $i$ would be $\lesssim 3^{\circ}$ for $\varepsilon
\leq 0.35$ (within our assumed error $\delta_{i}$) and $\lesssim 6^{\circ}$ for $0.35
< \varepsilon < 0.55$ (within 2$\delta_{i}$). In our sample of Sphs, there are no
objects with $\varepsilon > 0.55$. Our results are listed in Tables \ref{tab:BCDs},
\ref{tab:Irrs}, and \ref{tab:Sphs}.

\begin{figure*}[tbp]
\centering
\includegraphics[angle=-90, width=17cm]{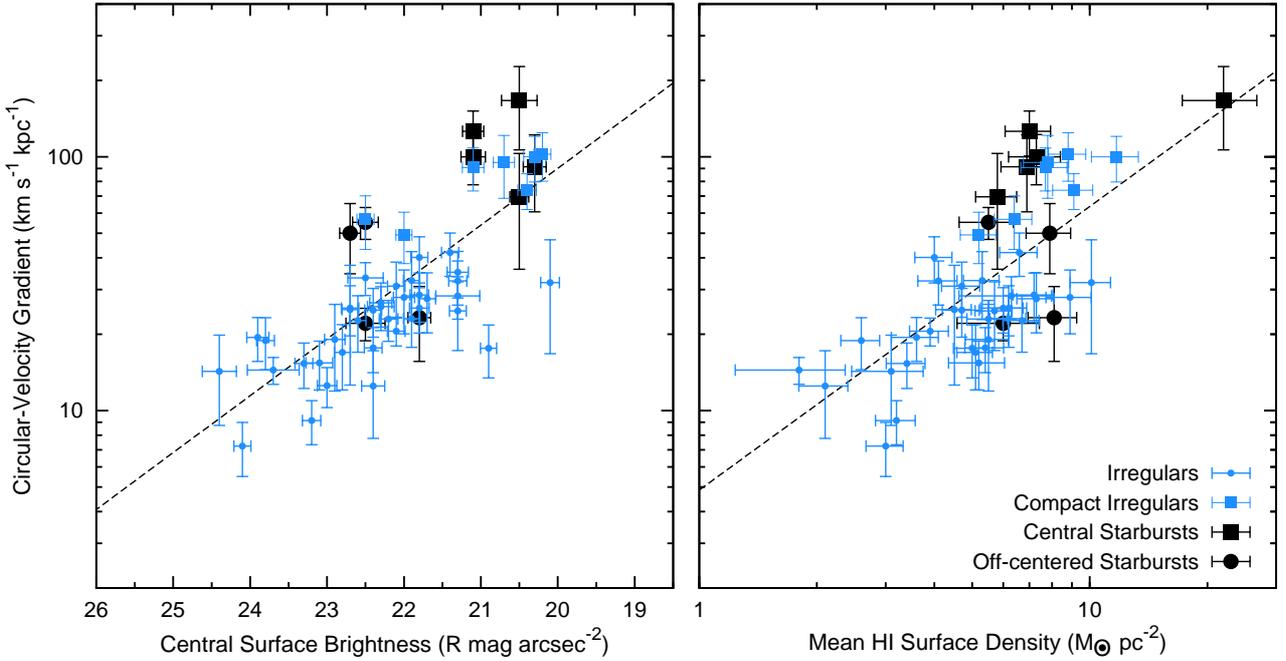}
\caption{\textit{Left:} the circular-velocity gradient $V_{R_{\rm d}}/R_{\rm d}$ versus the
$R$-band central surface brightness (inclination corrected). \textit{Right:} $V_{R_{\rm d}}/R_{\rm d}$
versus the mean \hi surface density within $R_{\rm opt}$. The black dots and squares show,
respectively, BCDs with a diffuse/off-centered starburst and BCDs with a centrally-concentrated
starburst. The blue dots and squares show, respectively, typical Irrs and compact Irrs (see
Sect.~\ref{sec:SFdwarf} for details). The dashed lines show a linear-fit to the data.}
\label{fig:GradPlot1}
\end{figure*}
\subsection{\label{sec:SFprop}Gas and star formation properties}

We collected information about the \hi content, ionized-gas metallicity, and star
formation for all the BCDs and Irrs in our sample. These are briefly described in
the following.

We calculated \hi masses using the standard formula
\begin{equation}
 M_{\hi}[M_{\odot}] = 236 \times D \, [{\rm Mpc}]^{2} \times S_{\hi} [\rm{mJy \, km \, s^{-1}}],
\end{equation}
where $S_{\hi}$ is the observed \hi flux from interferometric 21-cm line observations.
The error $\delta_{M_{\hi}}$ on $M_{\hi}$ is given by:
\begin{equation}
\delta_{M_{\hi}}^{2} = ( 472 \, D \, S_{\hi} \delta_{D} )^{2} + ( 236 \, D^{2} \, \delta_{S_{\hi}} )^{2}.
\end{equation}
Since the error on $S_{\hi}$ is usually not given in the original papers, we assumed
$\delta_{S_{\hi}} = 0.1 \, S_{\hi}$ (the typical calibration error of standard \hi
observations).

Following \citet{Swaters2002a}, we define the optical radius $R_{\rm opt} = 3.2 R_{\rm d}$.
$R_{\rm opt}$ is \textit{not} an isophotal radius and, thus, allows us to compare the sizes
of galaxies with different central surface brightnesses. For a HSB exponential disk with
$\mu_{0}(B) = 21.65$ mag \citep{Freeman1970}, $R_{\rm opt}= 3.2~ R_{\rm{d}}$ is equivalent
to the usual isophotal radius $R_{25}$. We define $\avghi$ as the mean \hi surface density
within $R_{\rm opt}$. $\avghi$ is derived from the observed \hi map using an elliptical
aperture, and is corrected for $i$ by multiplying the mean \hi column density $\overline{N}_{\hi}$
by $\cos(i)$. We calculated $\avghi$ for our sample of BCDs, while \citet{Swaters2002a}
provide $\avghi$ for 38 Irrs. We point out that $\avghi$ is distance-independent and
nearly unaffected by the linear resolution (in kpc) of the \hi observations, as the
galaxies in our sample are resolved within $R_{\rm opt}$. The error $\delta_{\avghi}$
on $\avghi$ is given by
\begin{equation}
\delta_{\avghi}^{2} = [\cos(i)\delta_{\overline{N}_{\hi}}]^{2} + [\overline{N}_{\hi}\sin(i)\delta_{i}]^{2}
\end{equation}
where $\delta_{\overline{N}_{\hi}}$ is assumed to be 10$\%$ of $\overline{N}_{\hi}$.

We calculated SFRs using the H$\alpha$ luminosities $L_{\rm{H}\alpha}$ from
\citet{Kennicutt2008}, scaled to our assumed distances, and the standard
calibration from \citet{Kennicutt1998}. This calibration assumes a Salpeter
initial mass function (IMF) from 0.1 to 100 M$_{\odot}$ and solar metallicity.
The latter assumption is clearly not valid for dwarf galaxies, which generally
have sub-solar metallicities (see Tables \ref{tab:BCDs} and \ref{tab:Irrs2}).
According to \citet{Lee2009b}, the ratio SFR/L$_{\rm{H}\alpha}$ for a galaxy with
$Z \simeq Z_{\odot}/5$ is a factor of $\sim$0.7 lower with respect to a galaxy
with $Z = Z_{\odot}$, thus our SFRs may be slightly overestimated. Other common
calibrations, which assume a Kroupa IMF and different stellar evolutionary
models, would give SFRs that are lower by a factor of $\sim$0.68 \citep[see]
[]{Kennicutt2012}. \citet{Kennicutt2008} provides $L_{\rm{H}\alpha}$ for
8 BCDs and 29 Irrs present in our sample. For the remaining BCD (I~Zw~18),
we used the H$\alpha$ luminosity from \citet{GilDePaz2003}. For another 8
Irrs, we used the H$\alpha$ SFRs calculated by \citet{James2004} (scaled
to our assumed distances and uncorrected for internal extiction), who also
used the \citet{Kennicutt1998} calibration. Besides the uncertainties on
the absolute SFR calibration, the errors on the SFRs are of the order of
$2 \delta_{D}/D$.

We then calculated the SFR surface density $\Sigma_{\rm SFR}= \rm{SFR}
/(\pi R_{\rm{opt}}^{2})$ and the ratio SFR/$M_{\rm bar}$, which is a
baryonic version of the specific SFR ($\rm{sSFR=SFR}/M_{*}$). The baryonic
mass (stars and atomic gas) was estimated from the baryonic TF relation,
calibrated by \citet{McGaugh2012} as $M_{\rm bar} [M_{\odot}]
= 47 \times V_{\rm circ}^{4} [\rm{km\,s^{-1}}]$ with an accuracy of
$\sim$10$\%$. For the 16 Irrs and 2 BCDs that do not reach the flat part
of the rotation curve, this baryonic mass may be slightly underestimated.
We also calculated two types of gas-depletion times: i) $\tau_{\rm global}
= 1.33 \, M_{\hi}/\rm{SFR}$, which considers the total atomic gas mass,
and ii) $\tau_{\rm local} = 1.33\, \avghi/\Sigma_{\rm SFR}$, which considers
only the atomic gas mass inside $R_{\rm opt}$. The factor 1.33 takes
into account the contribution of Helium \citep[e.g.][]{deBlok2008}. Note
that $\Sigma_{\rm SFR}$, $\tau_{\rm global}$, and $\tau_{\rm local}$ are
distance independent; the errors depend on the accuracy of the \hi and H$\alpha$
flux calibrations (typically $\sim$10$\%$) and on the SFR calibration. Since
the SFRs may be \textit{overestimated} up to a factor of $\sim$2 due to the
assumptions on the IMF and metallicity, the gas depletion times may be slightly
\textit{underestimated}. Finally, we compiled H$\alpha$+\NII equivalent
widths (EW) and ionized-gas metallicities $12+\log(O/H)$ from the literature.
Our results are listed in Tables \ref{tab:BCDs2} and \ref{tab:Irrs2}.

\section{\label{sec:res}Results}

\subsection{\label{sec:SFdwarf}Gas-rich dwarf galaxies}

We start by comparing the dynamical properties of gas-rich, star-forming
dwarfs (BCDs and Irrs). As discussed in Sect.~\ref{sec:V/R}, for
a bulgeless galaxy with a nearly exponential luminosity profile, 
$V_{R_{\rm d}}/R_{\rm d}$ is a good proxy for the circular-velocity
gradient $d_{R}V(0) \propto \sqrt{\rho_{0}}$, where $\rho_{0}$ is the central
dynamical mass density (including gas, stars, and dark matter). In
Fig.~\ref{fig:GradPlot1}, $V_{R_{\rm d}}/R_{\rm d}$ is plotted versus
the central surface brightness $\mu^{i}_{0}$ (left) and the mean \hi
surface density $\avghi$ (right). $V_{R_{\rm d}}/R_{\rm d}$ correlates
with both $\mu^{i}_{0}$ and $\avghi$ \citep[see also][]{Lelli2012b,
Lelli2013}. Gas-rich dwarfs with a high central dynamical mass density
(high $V_{R_{\rm d}}/R_{\rm d}$) have also a high central surface
brightness and a high \hi surface density within the stellar body.
To quantify the statistical significance of these relations, we
calculated the Pearson's correlation coefficient $\rho_{\rm cc}$,
where $\rho_{\rm cc} = \pm 1$ for an ideal linear correlation/anticorrelation 
while $\rho_{\rm cc} = 0$ if no correlation is present. We found
that both correlations are highly significant: the $V_{R_{\rm d}}
/R_{\rm d} - \mu^{i}_{0}$ diagram has $\rho_{\rm cc}\simeq-0.8$,
whereas the $V_{R_{\rm d}}/R_{\rm d}-\avghi$ diagram has
$\rho_{\rm cc} \simeq 0.7$. A linear, error-weighted fit
to the data returns
\begin{equation}\label{eq:GradvsSB}
\log(V_{R_{\rm d}}/R_{\rm d}) = (-0.22 \pm 0.03) \, \mu^{i}_{0} + (6.4 \pm 0.6),
\end{equation}
and
\begin{equation}\label{eq:GradvsHI}
\log(V_{R_{\rm d}}/R_{\rm d}) = (1.1 \pm 0.2) \, \log(\avghi) + (0.7 \pm 0.1).
\end{equation}
The left panel of Fig.~\ref{fig:GradPlot1} is nearly equivalent to the
lower part of the scaling relation described in \citet{Lelli2013}, which
holds for both irregular and spiral galaxies and extends for 2 orders of
magnitude in $d_{R}V(0)$ and 4 orders of magnitude in surface brightness.
The values of the slope and intersect in Eq.~\ref{eq:GradvsSB} are in
close agreement with those found in \citet{Lelli2013} ($-0.22 \pm 0.02$
and $6.3 \pm 0.4$, respectively).

\begin{figure*}[tbp]
\centering
\includegraphics[angle=-90, width=18cm]{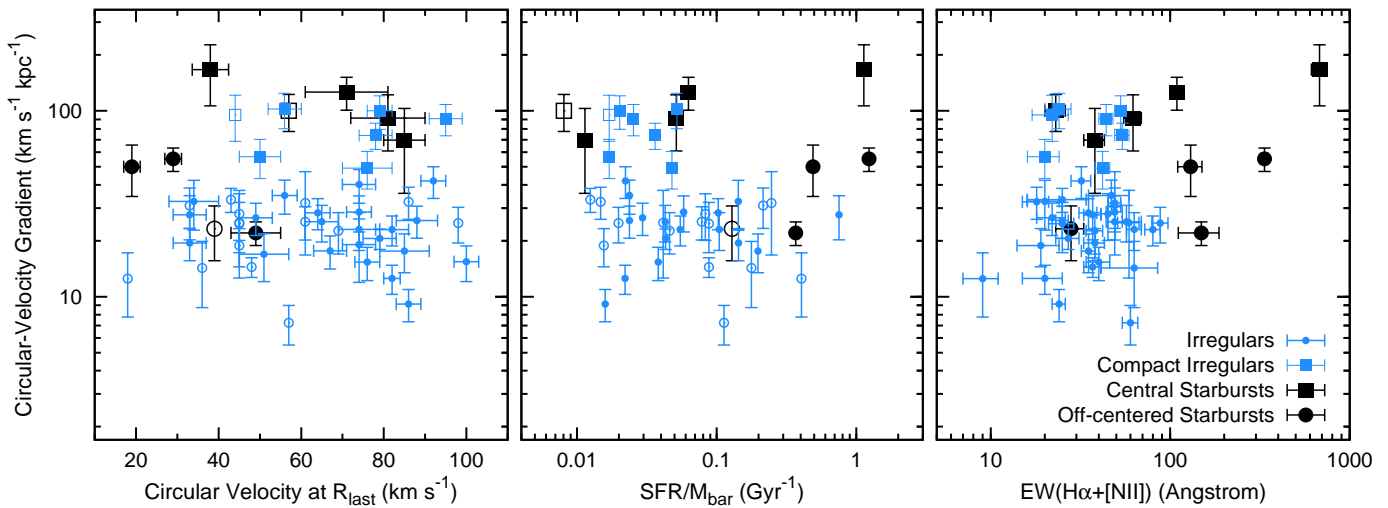}
\caption{\textit{Left:} $V_{R_{\rm d}}/R_{\rm d}$ versus the circular velocity
at the last measured point $V_{\rm last}$. \textit{Middle:} $V_{R_{\rm d}}/R_{\rm d}$
versus SFR/$M_{\rm bar}$, where $M_{\rm bar}$ is estimated using the baryonic
TF relation. \textit{Right:} $V_{\rm{R_{\rm d}}}/R_{\rm d}$ versus the
H$\alpha$+\NII equivalent width. Symbols are the same as in Fig.~\ref{fig:GradPlot1},
except for the left and middle panels, where open symbols indicate galaxies
with rotation curves that keep rising at $V_{\rm last}$, thus their value
of SFR/$M_{\rm bar}$ may be overestimated.}
\label{fig:GradPlot2}
\end{figure*}

The previous correlations are completely driven by the \textit{local},
inner properties of the galaxies ($\mu^{i}_{0}$ and $\avghi$) and not
by \textit{global} properties, such as the total baryonic mass or the
total dynamical mass. This is illustrated in Fig.~\ref{fig:GradPlot2}
(left), where $V_{R_{\rm d}}/R_{\rm d}$ is plotted versus the circular-velocity
at the last measured point $V_{\rm last}$. Clearly, there is no correlation
($\rho_{\rm cc} \simeq 0.01$). Similarly, we found no correlation with
the absolute magnitude $M_{\rm R}$ and with the dynamical mass (calculated
at the last measured point). At every value of $V_{\rm last}$ (or $M_{\rm R}$),
one can find both high-surface-brightness (HSB) dwarfs with a steeply-rising
rotation curve and low-surface-brightness (LSB) dwarfs with a slowly-rising
rotation curve (cf. with Fig.~\ref{fig:GradPlot1}, left). In particular,
for a given $V_{\rm last}$, BCDs typically have higher values of
$V_{R_{\rm d}}/R_{\rm d}$ than the bulk of Irrs.

Fig.~\ref{fig:GradPlot1} clearly shows that most BCDs are in the top-right
part of the $V_{R_{\rm d}}/R_{\rm d}-\mu^{i}_{0}$ and $V_{R_{\rm d}}
/R_{\rm d}-\overline{\Sigma}_{\hi}$ diagrams. This suggests that the
starburst activity is closely linked to the inner shape of the potential
well and to the central gas surface density. We can distinguish,
however, between two types of BCDs:
i) centrally-concentrated starbursts (NGC~1705, NGC~2915, NGC~4214,
NGC~6789, and I~Zw~18), and ii) diffuse and/or off-centered starbursts
(NGC~2366, NGC~4068, UGC~4483, and I~Zw~36). The former (black squares)
have the highest values of $V_{R_{\rm d}}/R_{\rm d}$ and $\mu^{i}_{0}$,
whereas the latter (black dots) show moderate values of $V_{R_{\rm d}}
/R_{\rm d}$ and $\mu^{i}_{0}$. In particular, NGC~2366 and UGC~4483 are
prototype ``cometary'' BCDs \citep[e.g.][]{Noeske2000}, as the starburst
region is located at the edge of an elongated LSB stellar body (see the
Atlas in Lelli et al., submitted). I~Zw~36 has an off-centered starburst
region superimposed on an elliptical stellar body, and may be a ``cometary-like''
BCD observed close to face-on. Finally, NGC~4068 has several small
star-forming regions spread over the entire stellar body. It is possible
that BCDs with a diffuse/off-centered starburst are different from BCDs
with a centrally-concentrated starburst in terms of their structure and
dynamics. This may be related to different evolutionary histories and/or
triggering mechanisms.

Several Irrs have circular-velocity gradients comparable to those of BCDs
($\gtrsim$ 45~km~s$^{-1}$~kpc$^{-1}$). These objects, shown by blue squares
in Figs \ref{fig:GradPlot1} and \ref{fig:GradPlot2}, are the following (the
classification from \citealt{RC3} is given): UGC~3711 (IBm), UGC~3966 (Im),
UGC~5721 (SBd?), UGC~7232 (Im pec), UGC~7261 (SBdm), UGC~7690 (Im), and
UGC~8490 (Sm). These galaxies have structural and dynamical properties more
similar to BCDs than to typical Irrs. In particular, they have HSB exponential
profiles with relatively-small scale-lengths ($\lesssim$1~kpc, see the
right panel of Fig.~\ref{fig:Sph}) and strong concentrations of gas near
the galaxy center (cf.~\citealt{Swaters2002b}). This suggests that either
they are also starbursting dwarfs (and may be considered as BCDs) or
they are progenitors/descendants of BCDs. Except for the barred galaxies
UGC~3711 and UGC~7261, the surface brightness profiles of these Irrs do
\textit{not} show the central ``light excess'' that is typically observed
in BCDs \citep[cf.][]{Swaters2002a}, thus it is likely that they are
\textit{not} experiencing a starburst at the present epoch. We will
refer to them as compact Irrs. The study of the SFHs of these compact
Irrs may be crucial to address their relation to BCDs.

\subsection{Gravitational potential and starburst indicators}

To clarify the relation between the gravitational potential and the
star-formation, we plotted $V_{R_{\rm d}}/R_{\rm d}$ against several
starburst indicators: the ratio SFR/$M_{\rm bar}$ (similar to the
$\rm{sSFR=SFR}/M_{*}$), the equivalent width EW(H$\alpha$+\NIIA), the
SFR surface density $\Sigma_{\rm SFR}$, and the gas depletion times
$t_{\rm local}$ and $t_{\rm global}$ (see Sect.~\ref{sec:SFprop}
for details). In the literature, there is no general agreement about
which of these indicators best identifies a starburst galaxy.

\begin{figure*}[tbp]
\centering
\includegraphics[angle=-90, width=18cm]{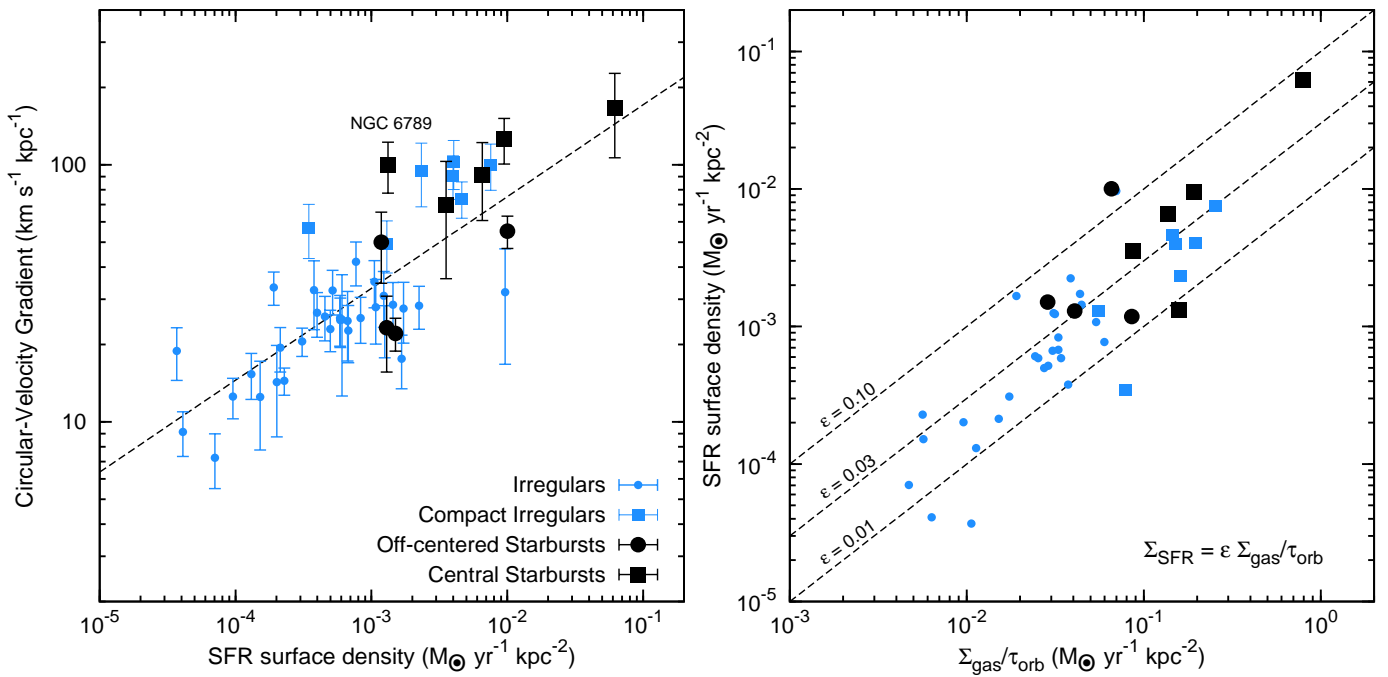}
\caption{\textit{Left:} $V_{R_{\rm d}}/R_{\rm d}$ versus the SFR surface
density $\Sigma_{\rm SFR}$. The dashed line shows a linear-fit to the data.
The position of the post-starburst dwarf galaxy NGC~6789 is indicated.
\textit{Right:} $\Sigma_{\rm SFR}$ versus $\Sigma_{\rm gas}/\tau_{\rm orb}$,
where $\Sigma_{\rm gas}$ considers the atomic gas only and $\tau_{\rm orb}$
is the orbital time on the solid-body portion of the rotation curve.
The dashed lines show a linear relation for different value of $\epsilon$.
Symbols are the same as in Fig.~\ref{fig:GradPlot1}.}
\label{fig:GradvsSFRdens}
\end{figure*}
The ratio SFR/$M_{\rm bar}$ quantifies the star-formation activity of
a galaxy with respect to its baryonic mass (stars and atomic gas). This
is comparable to SFR/$M_{*}$, which is often used for massive galaxies.
We prefer to use SFR/$M_{\rm bar}$ for two reasons: i) we do not have a
direct estimate of $M_{*}$ for the Irrs, whereas we can estimate $M_{\rm bar}$
using the baryonic TF relation (see Sect.~\ref{sec:SFprop} for details),
and ii) in some dwarf galaxies the atomic gas mass can be as high as
the stellar mass, thus a LSB Irr that has been inefficiently forming
stars during the Hubble time might have a relatively-high
SFR/$M_{*}$-ratio but a low SFR/$M_{\rm bar}$-ratio. The differences
between SFR/$M_{\rm bar}$ and SFR/$M_{*}$, however, are typically within
a factor of $\sim$2. Figure~\ref{fig:GradPlot2} (middle) shows that there
is no correlation between $V_{R_{\rm d}}/R_{\rm d}$ and SFR/$M_{\rm bar}$
($\rho_{\rm cc}\simeq-0.1$). Moreover, BCDs and Irrs can have similar values
of SFR/$M_{\rm bar}$, suggesting that this is not a good starburst indicator.

The EW(H$\alpha$) is thought to trace the stellar birthrate parameter $b$,
defined as the ratio of the current SFR to the past, average SFR over the
galaxy lifetime \citep[e.g.][]{Kennicutt1998}. \citet{Lee2009a} argued
that an EW(H$\alpha$)$\gtrsim100~\mathring{\rm{A}}$ corresponds to
$b \gtrsim 2.5$ and, thus, identifies a starburst. Fig.~\ref{fig:GradPlot2}
(right) shows that there is no strong correlation between EW(H$\alpha$+\NIIA)
and $V_{R_{\rm d}}/R_{\rm d}$ ($\rho_{\rm cc}\simeq0.3$). Most BCDs, as
expected, have very high values of EW(H$\alpha$+\NIIA), although some
of them have EW(H$\alpha$+\NIIA)$<$100 $\mathring{\rm A}$. As pointed out
by \citet{McQuinn2010a}, starbursts are events lasting for a few 100 Myr,
whereas the H$\alpha$ emission probes the star-formation activity over
shorter timescales ($\lesssim$10~Myr), thus fluctuations in the SFH over
a few Myr may explain why a EW(H$\alpha$) treshold misidentifies some BCDs.

The $\Sigma_{\rm SFR}$ normalizes the SFR by the area of the stellar body.
Fig.~\ref{fig:GradvsSFRdens} (left) shows that $V_{R_{\rm d}}/R_{\rm d}$
correlates with $\Sigma_{\rm SFR}$ ($\rho_{\rm cc} \simeq 0.8$). This
suggests that there is a close link between the star-formation activity
and the inner steepness of the potential well. A linear, error-weighted
fit to the data returns
\begin{equation}\label{eq:SFRdens}
\log(V_{R_{\rm d}}/R_{\rm d}) = (0.36 \pm 0.04) \, \log(\Sigma_{\rm SFR}) + (2.6 \pm 0.1).\\
\end{equation}
This correlation is expected from Fig.~\ref{fig:GradPlot1} (right) and the
Kennicutt-Schmidt (KS) law \citep[e.g.][]{Kennicutt1998b}. By combining
Eq.~\ref{eq:GradvsHI} and Eq.~\ref{eq:SFRdens}, indeed, one can obtain a
KS type of relation, that considers the atomic gas only and has a slope
of $\sim$3. This is in line with the results of \citet{Roychowdhury2009},
who investigated the KS law in 23 extremly-faint dwarf galaxies ($M_{\rm B}
\simeq-13$ mag) and found a slope of $\sim$2.5 by considering atomic gas
only and UV-based SFRs. As expected, most starbursting dwarfs are in the top-right
part of the $V_{R_{\rm d}}/R_{\rm d}-\Sigma_{\rm SFR}$ diagram. In BCDs,
the starburst typically increases the SFR by a factor of $\sim$5 to $\sim$10
\citep[e.g.][]{McQuinn2010a}. NGC~6789, which is the only \textit{known}
post-starburst galaxy in our sample (see \citealt{McQuinn2010a} and
Sect.~\ref{sec:disc2}), shows a large horizontal deviation with respect
to the main relation ($\gtrsim$1 dex). Intriguingly, compact Irrs
also have high values of $\Sigma_{\rm SFR}$, but they systematically
lie on the left side of the linear-fit by $\sim$0.5 to $\sim$1 dex,
thus they are consistent with being progenitors/descendants of BCDs.
However, the overall, large scatter on the $V_{R_{\rm d}}/R_{\rm d}
-\Sigma_{\rm SFR}$ relation prevents us from reaching any firm
conclusion about the nature of compact Irrs and their link with BCDs.

We point out that $V_{R_{\rm d}}/R_{\rm d} = 2\pi/\tau_{\rm orb}$,
where $\tau_{\rm orb}$ is the orbital time on the solid-body
portion of the rotation curve. \citet{Kennicutt1998b} found that,
for spiral galaxies and massive starbursts, $\Sigma_{\rm SFR}$
also correlates with $\Sigma_{\rm gas}/\tau_{\rm orb}$, where
$\tau_{\rm orb}$ was calculated at the outer edge of the optical
disk (presumably along the flat part of the rotation curve) and 
$\Sigma_{\rm gas}$ includes both atomic and molecular gas.
This correlation may be interpreted as the effect of spiral arms
triggering the star-formation \citep[e.g.][]{Kennicutt1998b}.
Fig.~\ref{fig:GradvsSFRdens} (right) shows that a similar
correlation (with $\rho_{\rm cc} \simeq 0.9$) holds also for
gas-rich dwarfs, in which the effect of density waves clearly
cannot be important. Here $\tau_{\rm orb}$ is calculated on the
solid-body portion of the rotation curve, while $\Sigma_{\rm gas}$
includes the atomic gas component only, since the molecular content
of low-metallicity, dwarf galaxies is very uncertain \citep[e.g.]
[]{Taylor1998, Leroy2008}. As discussed by \citet{Kennicutt1998b},
one might expect a linear relation of the form:
\begin{equation}\label{eq:torb}
\Sigma_{\rm SFR} = \epsilon \, \Sigma_{\rm gas}/\tau_{\rm orb}\\
\end{equation}
where $\epsilon$ is the fraction of gas that is converted into
stars during every orbit. For spiral galaxies and massive starbursts,
\citet{Kennicutt1998b} found that $\sim$10$\%$ of the available gas
(atomic plus molecular) is converted into stars during every orbit.
The dashed lines in Fig.~\ref{fig:GradvsSFRdens} (right) show
fractions $\epsilon = 0.01$, 0.03, and 0.10. Most dwarf galaxies
have $0.01\lesssim \epsilon\lesssim0.03$, but several BCDs seem
to have $\gtrsim0.03$, suggesting that they might be converting
gas into stars more efficiently than other gas-rich dwarfs.
However, metallicity and/or internal extinction may affect
the relative values of $\Sigma_{\rm SFR}$ in different galaxies,
thus it is unclear whether the differences in $\epsilon$ are
real \textit{or} due to the use of the same SFR calibration
for all the galaxies, without considering the possible effects
of internal extinction and metallicity.

\begin{figure*}[tbp]
\centering
\includegraphics[angle=-90, width=17cm]{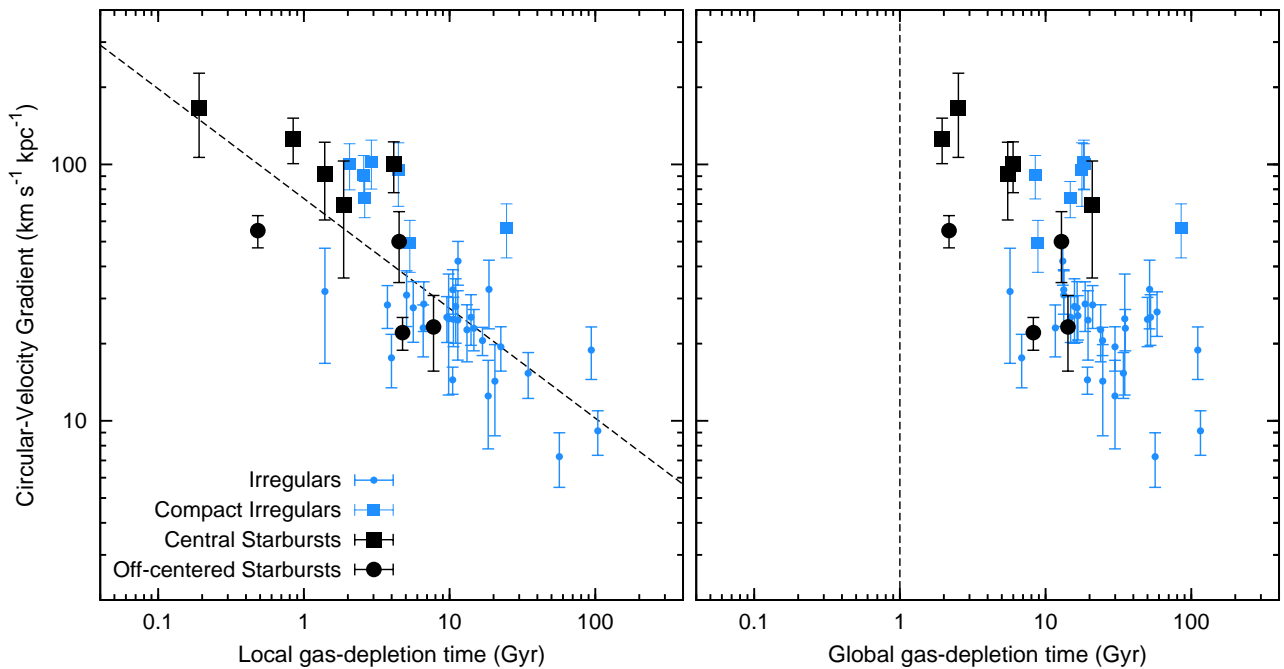}
\caption{\textit{Left:} $V_{R_{\rm d}}/R_{\rm d }$ versus the local
gas-depletion time, that takes into account the atomic gas mass within
the $R_{\rm opt}$. The dashed line show a linear fit to the data.
\textit{Right:} $V_{\rm{R_{\rm{d}}}}/R_{\rm{d}}$ versus the total
gas-depletion time, that takes into account the total atomic gas mass.
The dashed line shows the maximum duration of a typical starburst
($\sim$1~Gyr). Symbols are the same as in Fig.~\ref{fig:GradPlot1}.}
\label{fig:GradvsTdep}
\end{figure*}
Finally, $\tau_{\rm local}$ and $\tau_{\rm global}$ estimate the
time that a galaxy will take to consume its gas reservoir, if it
keeps forming stars at the current rate. $\tau_{\rm local}$ takes
into account only the atomic gas within the stellar component, while
$\tau_{\rm global}$ takes into account the total atomic gas. Since
our SFRs may be slightly overestimated, the values of $\tau_{\rm
local}$ and $\tau_{\rm global}$ may be underestimated by a factor of
$\sim$2 (see Sect.~\ref{sec:SFprop}). Several studies \citep[e.g.][]
{vanZee2001b, Hunter2004} have shown that gas-rich dwarfs have
long gas-depletion times ($>10$~Gyrs) and, thus, could keep forming
stars for several Hubble times. In Fig.~\ref{fig:GradvsTdep} (left),
we show that $V_{R_{\rm d}}/R_{\rm d}$ correlates with $\tau_{\rm local}$
($\rho_{\rm cc} \simeq -0.8$), as expected from Figs.~\ref{fig:GradPlot1}
(right) and \ref{fig:GradvsSFRdens} (left). BCDs and compact Irrs
typically have $\tau_{\rm local} \lesssim 5$~Gyr. A linear,
error-weighted fit to the data yields
\begin{equation}{\label{eq:tloc}}
\log(V_{R_{\rm{d}}}/R_{\rm{d}}) = (-0.43 \pm 0.06) \, \log(\tau_{\rm{local}}) + (1.87 \pm 0.06).\\
\end{equation}
Figure \ref{fig:GradvsTdep} (right), instead, shows that the relation
between $V_{R_{\rm d}}/R_{\rm d}$ and $\tau_{\rm global}$ is less well
defined ($\rho_{\rm cc} \simeq -0.6$). It also shows that $\tau_{\rm global}$
is significantly larger than $\tau_{\rm local}$, implying that Irrs and
BCDs have massive gas reservoirs outside the stellar component. For most
BCDs, both $\tau_{\rm local}$ and $\tau_{\rm global}$ are larger than the
typical durations of the burst (few 100 Myr), implying that they do not consume
their entire gas reservoir during the current event of intense star-formation.
Burst durations can be estimated using the SFHs derived by fitting the
color-magnitude diagrams of the resolved stellar populations \citep[e.g.]
[]{McQuinn2010a}. In particular, the sample of \citet{McQuinn2010a}
includes five ``fossil'' starbursting dwarfs, that allows us to estimate
the \textit{total} duration of the burst. If one defines the burst
duration as the period when the birthrate parameter $b > 2$,
Fig.~2 of \citet{McQuinn2010b} shows that the longest burst duration
is $\sim$850~Myr in UGC~9128. \citet{McQuinn2010b} report slightly
higher values as they use a less conservative definition of ``burst
duration''. We assume a fiducial value of 1~Gyr as the maximum duration
of a starburst; this is indicated in Fig.~\ref{fig:GradvsTdep} by a
verical dashed-line. All the BCDs in our sample have $\tau_{\rm global}
> 1$~Gyr, and most of them also have $\tau_{\rm local} > 1$~Gyr.

\subsection{\label{sec:Sphs}Gas-poor dwarf galaxies}

We now compare the structural and dynamical properties of
gas-rich dwarfs with those of rotating Sphs in the Virgo cluster
\citep{vanZee2004a, vanZee2004b}. These gas-poor dwarfs are
at higher distances than most gas-rich dwarfs in our sample;
beam-smearing effects, however, are not a serious issue in
the derivation of the stellar rotation curves, given that the
long-slit optical observations have significantly higher
angular resolution ($\sim$2$''$) than the \hi observations
($\sim$10$''$ to 30$''$). The stellar rotation velocities
have been corrected for pressure-support as described in
Appendix~\ref{sec:asymDrift}.

Figure \ref{fig:Sph} (left) shows that rotating Sphs follow the same
correlation between $V_{R_{\rm d}}/R_{\rm{d}}$ and $\mu^{i}_{0}$
defined by Irrs and BCDs. Moreover, these rotating Sphs have values
of $V_{R_{\rm d}}/R_{\rm d}$ and $\mu^{i}_{0}$ comparable with those
of BCDs and compact Irrs. In Fig.~\ref{fig:Sph} (right), we plot
$R_{\rm d}$ against $\mu^{i}_{0}$; the dashed lines correspond to
exponential profiles with a fixed total magnitude. The rotating
Sphs have total $R$-band magnitudes in the range $-16 \lesssim
M_{\rm R} \lesssim -18$, comparable to the Irrs considered here.
In general, for a given $M_{\rm R}$, the values of $\mu^{i}_{0}$
and $R_{\rm d}$ of rotating Sphs are, respectively, higher and
smaller than those of typical Irrs, but comparable with those of
some BCDs and compact Irrs. Therefore, the structural and dynamical
properties of rotating Sphs in the Virgo cluster appear similar
to those of BCDs and compact Irrs in the field and nearby groups.

Since our sample of gas-poor dwarfs is relatively small, it is
unclear whether rotating Sphs are necessarily more compact than
typical Irrs, or whether this is the result of selection effects.
It is clear, however, that a close link between the central
dynamical mass density ($V_{R_{\rm d}}/R_{\rm d}$) and the
stellar surface density ($\mu^{i}_{0}$) is present in any
kind of rotating galaxy \citep[see also][]{Lelli2013}.

\begin{figure*}[tbp]
\centering
\includegraphics[angle=-90, width=18cm]{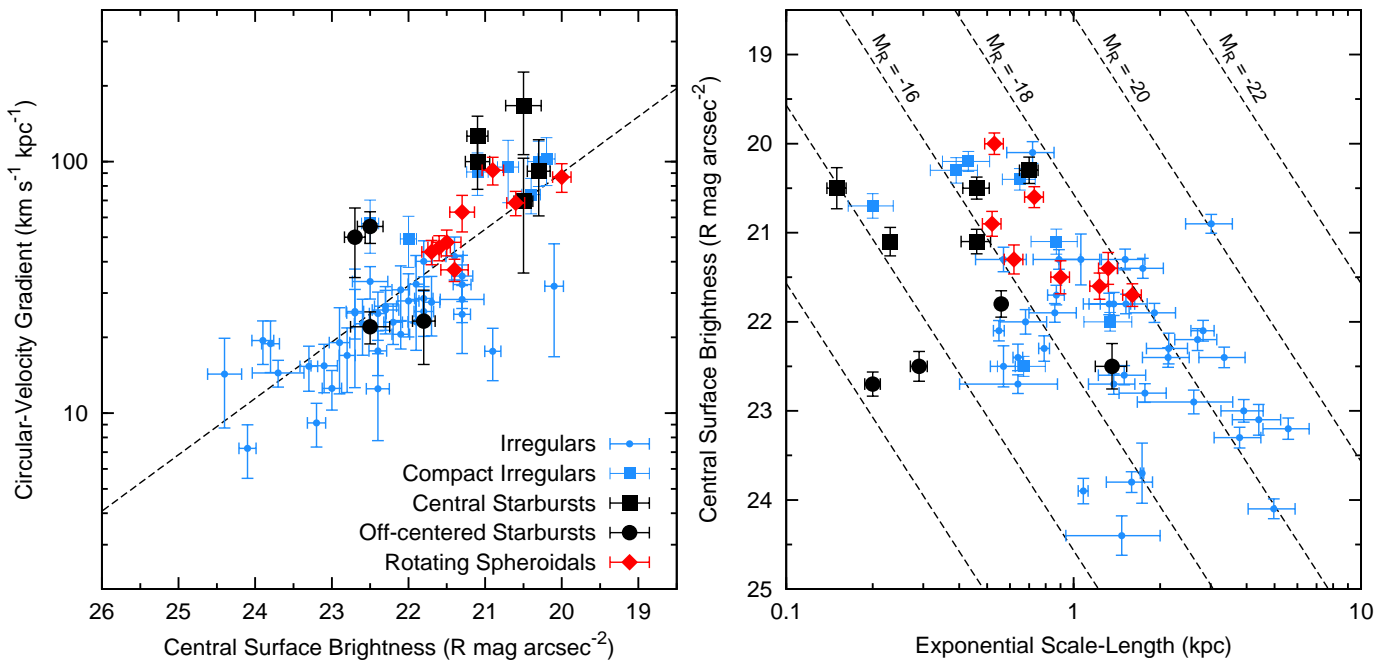}
\caption{\textit{Left:} $V_{R_{\rm d}}/R_{\rm d}$ versus $\mu^{i}_{0}$
including rotating Sphs in the Virgo cluster (red diamonds). The other
symbols are the same as in Fig.~\ref{fig:GradPlot1}. The dashed line shows
a linear fit to the Irr and BCD data (same as in Fig.~\ref{fig:GradPlot1}).
The stellar rotation velocities have been corrected for asymmetric drift
(see Appendix \ref{sec:asymDrift}). \textit{Right}: $R_{\rm d}$ against
$\mu^{i}_{0}$. The dashed lines correspond to exponential profiles with
a fixed total magnitude ($-22$, $-20$, $-18$, $-16$, $-14$, and $-12$
R mag).}
\label{fig:Sph}
\end{figure*}
\section{\label{sec:disc}Discussion}

\subsection{The relation between gravitational potential, gas surface
density, and starburst activity \label{sec:disc1}}

In Sect.~\ref{sec:V/R} we estimated the circular-velocity gradient
$d_{R}V(0) \simeq V_{R_{\rm d}}/R_{\rm d}$ for a sample of 60 dwarf
galaxies, including BCDs, Irrs, and Sphs. $V_{R_{\rm d}}/R_{\rm d}$
is a direct measure of the central dynamical mass density in a galaxy
(including gas, stars, and dark matter). In Sect.~\ref{sec:SFdwarf}
we showed that, for gas-rich dwarfs, $V_{R_{\rm d}}/R_{\rm d}$
correlates with i) the central surface brightness, ii) the mean
\hi surface density over the stellar body, and iii) the SFR surface
density. Starbursting dwarfs are in the upper parts of these
relations, as they have high \hi surface densities, high central
surface brightnesses, and high circular-velocity gradients,
indicating that they have a high central dynamical mass density.
This implies that the starburst activity is closely related to
the inner steepness of the potential well and the gas surface
density. The nature of this connection is unclear. There are,
however, two likely possibilities: i) the progenitors of BCDs are
unusually compact Irrs with a steep potential well, or ii) there
is a mechanism that concentrates the mass (gas, stars, and dark
matter) in typical Irrs, eventually causing a starburst. These
possibilities are discussed in the following.

If the progenitors of BCDs are compact Irrs, the high circular-velocity
gradient would imply high values of the critical surface-density
threshold for gravitational instabilities $\Sigma_{\rm crit}$
\citep{Toomre1964, Kennicutt1989} and make the gaseous disk
relatively stable against large-scale perturbations. For a 
self-gravitating gaseous disk, $\Sigma_{\rm crit}$ is given by
\begin{equation}\label{eq:Scrit}
 \Sigma_{\rm crit} = \alpha \dfrac{\sigma_{\rm gas} \kappa}{\pi G}
\end{equation}
where $\alpha$ is a dimensionless parameter (near unity) that
accounts for the finite thickness of the disk \citep[e.g.][]
{Martin2001}, $\sigma_{\rm gas}$ is the gas velocity dispersion,
and $\kappa$ is the epicyclic frequency given by
\begin{equation}
 \kappa = 1.41 \dfrac{V_{\rm circ}(R)}{R}\sqrt{1 + \dfrac{d\log V_{\rm circ}(R)}{d\log R}}.
\end{equation}
The possible role of $\Sigma_{\rm crit}$ in BCDs has been discussed
by \citet{Meurer1998} and \citet{vanZee2001}, who argued that, in
these compact galaxies, the gas could pile up in the center and reach
high surface densities, while the star formation would be delayed
until the critical surface-density treshold is reached, leading to
a starburst. We note that this should strictly apply only to the inner
regions of the galaxy ($R \lesssim 2 R_{\rm d}$), where the shape
of the rotation curve is close to a solid body and, thus, $\Sigma_{\rm crit}
\propto V_{R_{\rm d}}/R_{\rm d}$ (assuming that $\sigma_{\rm gas}$
and $\alpha$ are nearly constant with radius). In the outer regions
($R \gtrsim 2 R_{\rm d}$), instead, the rotation curve becomes flat
and $\Sigma_{\rm crit} \propto 1/R$, hence the gaseous disk might be
less stable. Thus, this picture does not require a ``bursty'' SFH
for the whole galaxy, i.e. strong bursts separated by quiescent
periods, as the star-formation may continue in the outer parts of
the galaxy. Star-formation at large radii is actually observed in
several Irrs and BCDs \citep[e.g.][]{Hunter2004}.

Toomre's model for gravitational instabilities may also provide a simple
explanation for the correlation between $V_{R_{\rm d}}/R_{\rm d}$ and
$\avghi$ (Fig.~\ref{fig:GradPlot1}, right). If $\Sigma_{\rm gas} \simeq
\Sigma_{\rm crit} \propto V_{R_{\rm d}}/R_{\rm d}$ and the product
$\alpha \times \sigma_{\rm gas}$ is approximately the same in every
dwarf galaxy, we would expect the following relation:
\begin{equation}\label{eq:GradCrit}
 \log \bigg(\dfrac{V_{R_{\rm d}}}{R_{\rm d}}\bigg) = \log(\Sigma_{\rm gas}) + \log\bigg(\dfrac{\pi G}{\alpha \sigma_{\rm gas}}\bigg).
\end{equation}
Remarkably, the observed slope of the $V_{\rm R_{d}}/R_{\rm d} - \avghi$
relation is consistent with 1 within the uncertainties (see Eq.~\ref{eq:GradvsHI}).
Moreover, the value of the intersect (corrected for the presence of Helium)
imply $\alpha \times \sigma_{\rm gas}=3.6$ km~s$^{-1}$, in close agreement
with the value of $\sim$4 km~s$^{-1}$ found by \citet{Kennicutt1989} and
\citet{Martin2001} in the outer regions of spiral galaxies. Note that
Eq.~\ref{eq:Scrit} and Eq.~\ref{eq:GradCrit} are valid in the case of a
self-graviting gaseous disk. If one considers also the gravitational effect
of stars, the condition for the stability of the two-components disk is
more complex \citep[e.g.][]{Rafikov2001}, and it is unclear whether the
linear relation in Fig.~\ref{fig:GradPlot1} (right) may be simply explained.

There are several mechanisms that can cause a concentration of
mass in a galaxy, both internal (bar-like torques) and/or external
(interactions/mergers). First, we discuss internal mechanisms.
BCDs and compact Irrs are not necessarily barred, thus a bar-driven
inflow seems an unlikely general mechanisms. However, \citet{Hunter2004}
speculated that dwarf galaxies may have bars made of dark matter,
while \citet{Bekki2002} argued that rotating, triaxial dark-matter
haloes can exert large-scale torques and lead to mass inflow.
\citet{Elmegreen2012} proposed that massive clumps of gas and
young stars can exchange angular momentum with halo stars and
dark matter particles and, thus, may spiral inward to the galaxy
center, leading to a central starburst. This scenario would imply
an evolutionary trend from BCDs with an off-centered starbursts
(e.g. NGC~2366 and I~Zw~36) to BCDs with a centrally-concentrated
starburst (e.g. NGC~1705 and NGC~4214). Simulations of high-redshift
galaxies suggest that the clump-instability creates a central mass
concentration similar to a bulge, that is photometrically distinct from
the underlying exponential disk \citep[e.g.][]{Immeli2004, Bournaud2007}.
The compact Irrs identified in Sec.~\ref{sec:SFdwarf}, instead,
have surface brightness profiles described by a single HSB
exponential (with the exception of the barred galaxies UGC~3711
and UGC~7261; cf. \citealt{Swaters2002b}). BCDs often show surface
brightness profiles with an inner light excess, but these light
enhancements typically have very blue colors and do not appear
to trace the stellar mass \citep[e.g.][]{Papaderos1996}.

Regarding external mechanims, it is well established that galaxy
interactions and mergers can cause gas inflow and lead to a central
starburst \citep[e.g.][]{Barnes1996, Bekki2008}. Moreover, dwarf
galaxies are thought to be relatively fragile systems and external
perturbations may significantly change their mass distribution,
possibly leading to an overall contraction of the stellar and
gaseous disks \citep[e.g.][]{Hunter2004}. About half of the BCDs
in our sample have disturbed and asymmetric \hi morphologies in
their outer parts, suggesting either a recent interaction/merger
between gas-rich dwarfs \citep[see e.g. I~Zw~18 in][]{Lelli2012a}
or gas accretion from the environment \citep[e.g.][]{Keres2005,
Dekel2006}. Disturbed \hi morphologies have been found also in
other studies of BCDs \citep[e.g.][]{LopezSanchez2010b, Ashley2013}
and several interacting BCDs are known \citep[e.g.][]{Cox2001,
Ekta2006, Ekta2009}. Thus, the hypothesis that interactions/mergers
cause the central concentration of mass (gas, stars, and/or dark matter)
and trigger the starburst is in overall agreement with the observations.

\subsection{BCDs and the evolution of dwarf galaxies \label{sec:disc2}}

In the following, we discuss the possible evolutionary links between
BCDs, Irrs, and Sphs. The emerging picture is that a BCD in isolation
most likely evolves into a compact Irr, but environmental processes
(such as ram-pressure stripping) may transform compact Irrs and BCDs
into rotating Sphs.

\subsubsection{The link between BCDs and compact Irrs}

In Sect.~\ref{sec:SFdwarf} we pointed out that the global gas-depletion
times of BCDs are much larger than the typical durations of a starburst
($<$1 Gyr). This implies that BCDs generally do not consume their entire
gas reservoir. Moreover, both optical and X-rays observations suggest
that BCDs do not expell a large quantity of gas out of their potential
well (see also Lelli et al., submitted). Observations of the Na~D absorption
doublet have shown that ouflows are common in starbursting dwarfs, but the
velocities of the outflowing gas do \textit{not} clearly exceed the galaxy
escape velocity \citep[][]{Schwartz2004}. Similarly, studies of the H$\alpha$
kinematics have found that the warm gas is usually gravitationally-bound
to the galaxy \citep[e.g.][]{Martin1996, Martin1998, vanEymeren2009a,
vanEymeren2009b, vanEymeren2010}. X-ray studies have revealed that some
BCDs have diffuse coronae of hot gas at $T\simeq10^{6}$~K \citep{Ott2005a,
Ott2005b}, which are presumably associated with outflows but have very
low masses, only $\sim$1$\%$ of the \hi masses. Thus, a BCD \textit{in
isolation} most likely evolves into another gas-rich dwarf as the starburst fades.

In Sect.~\ref{sec:SFdwarf} we also identified several compact Irrs that
have structural and dynamical properties similar to BCDs. It is clear that,
unless a redistribution of mass takes place, the descendants of BCDs must
be compact Irrs. A strong redistribution of mass seems unlikely for two reasons:
i) in dwarf galaxies there are no \textit{known} internal mechanisms that
are able to redistribute stars and dark matter, such as radial migrations
due to bars and density waves \citep[e.g.][]{Minchev2011}; and ii) stellar
feedback seems unable to expel a large quantity of gas from the potential
well and, thus, cause a significant expansion of the stellar and dark matter
components. In particular, \citet{Papaderos1996} and \citet{GilDePaz2005}
showed that the scale-lengths of the old stellar component of BCDs should,
on average, increase by a factor of $\sim$2 to be consistent with those of
typical Irrs and Sphs. \citet{Papaderos1996} proposed a simple, spherically-symmetric
model to quantify the effect of ouflows on the evolution of the stellar
body of an isolated BCD; their equations 12 and 13 can be written as
\begin{equation}
 \dfrac{R^{\rm i}_{\rm d}}{R_{\rm d}^{\rm f}} = 1 - \dfrac{M_{\rm out}}{M^{\rm i}_{\rm bar}}f^{\rm i}_{\rm bar}
\end{equation}
where $R^{\rm i}_{\rm d}$ and $R_{\rm d}^{\rm f}$ are, respectively, the
scale-lengths in the initial (starburst) and final (post-starburst) stages
of the system, $M_{\rm out}$ is the gas mass that is ejected, $M^{i}_{\rm
bar} = M_{*} + M_{\rm gas}$ is the baryonic mass in the initial stage,
and $f^{\rm i}_{\rm bar} = M^{\rm i}_{\rm bar}/M^{\rm i}_{\rm dyn}$ is
the initial baryonic fraction within the stellar body. In Lelli et al.
(submitted), we showed that BCDs have, on average, $f_{\rm bar}\simeq 0.3$ to
0.5, depending on the assumptions on the IMF and on the molecular gas content.
In order to have $R^{\rm i}_{\rm d}/R_{\rm d}^{\rm f} = 1/2$, one would
need unphysical values of $M_{\rm out}/M^{\rm i}_{\rm bar}\simeq1$ to 1.7,
that would leave no baryons in the potential well. Thus, even if outflows
would expell a significant quantity of gas, they could not easily explain
the differences in the scale-lengths of BCDs and typical Irrs/Sphs.

In this evolutionary context, the BCD NGC~6789 deserves special attention.
NGC~6789 resides in the Local Void and is extremely isolated, as the
nearest massive galaxy (NGC 6946) is at a projected distance  of 2.5~Mpc
\citep{Drozdovsky2001}. The SFHs from \citet{McQuinn2010a} and
\citet{GarciaBenito2012} show that the starburst ended about $\sim$500~Myr
ago and the galaxy is now forming stars at a lower rate. Thus, NGC~6789 is
a post-starburst dwarf galaxy. The system still has a considerable gas
reservoir ($\sim$2$\times$10$^{7}$~M$_{\odot}$) and the gas depletion time
is long ($\sim$6 Gyr). The rotation curve rises steeply in the inner
parts, indicating that there is a strong central concentration of mass.
Thus, NGC~6789 did not consume its entire gas reservoir during the
starburst and did not experience a strong redistribution of mass in
the last $\sim$500~Myr, in agreement with our previous reasoning.

\subsubsection{The link between BCDs and rotating Sphs}

In Sect.~\ref{sec:Sphs}, we showed that rotating Sphs in the Virgo cluster
have central dynamical mass densities similar to those of BCDs and compact
Irrs. It is likely, therefore, that rotating Sphs are the descendants of
BCDs and compact Irrs, providing that some mechanism removes the gas from
the galaxy. As we already pointed out, supernova feedback seems unable to
entirely expell the ISM of a BCD, thus the best candidate is ram-pressure
stripping due to either the hot intra-cluster medium \citep[e.g.][]
{Kormendy2012} or the hot coronae that are thought to surround massive
galaxies \citep[e.g.][]{Gatto2013}. The rotating Sphs in our sample,
however, are still pressure-supported in the inner regions (with $v_{\rm rot}
/\sigma$ between $\sim$0.3 to $\sim$1), whereas the \textit{gaseous disks}
of BCDs and Irrs are rotation-supported (with $v_{\rm rot} /\sigma > 1$).
In particular, one may expect that the newly-formed stars retain the angular
momentum of the \hi disk, which is the site of the recent star-formation.
Thus, an additional mechanism that heats the stellar disk of a BCD/compact
Irr may be required for a morphological transformation into a rotating Sph.
Possible mechanisms are galaxy harassment by a galaxy cluster \citep[e.g.]
[]{Moore1998} or tidal stirring by a nearby massive companion \citep[e.g.]
[]{Mayer2006}. We warn, however, that the comparison between different
kinematical tracers (as stars and \hiA) may be misleading. \citet{Leaman2012}
studied the stellar kinematics of the Local Group irregular WLM and found
that its stellar disk has $v_{\rm rot}/\sigma \sim 1$, whereas the \hi
disk has $v_{\rm{rot}}/\sigma \sim 7$. Thus, the stellar and \hi kinematics
in a dwarf galaxy may be remarkably different, and the stellar component
of an Irr may be very similar to that of a Sph. In this case, the evolution
from a BCD/compact Irr to a rotating Sph would just require gas removal,
given that the current starburst typically creates only a small fraction
of the total stellar mass ($\sim$10$\%$, e.g. \citealt{McQuinn2010b}) and, thus,
the newly-formed stars cannot strongly affect the overall stellar kinematics.

The evolutionary sequence BCDs $\rightarrow$ compact Irrs $\rightarrow$ 
rotating Sphs may solve some long-standing issues about the direct
transformation of Irrs into \textit{bright} Sphs: i) Irrs are expected
to fade after the cessation of the star-formation and this would result
in central surface brightnesses that are significantly lower than those
of the brightest Sphs in galaxy clusters \citep{Bothun1986, Davies1988},
and ii) most bright Sphs show central nuclei whereas Irrs do not
\citep[e.g.][]{Cote2006, Lisker2007}. In fact, regardless of the
details of the triggering mechanism, the occurrence of a starburst
in an Irr should necessarily i) create a central concentration of
mass and increase the central surface brightness, and might possibly
ii) form a central nucleus by the inspiralling and merging of star
clusters \citep[][]{Gnedin2013}, which are often found in BCDs
\citep[e.g.][]{Annibali2009, Annibali2011}. In particular, the nuclei
of Sphs are generally bluer than the overall stellar body \citep[e.g.]
[]{Lotz2004}, suggesting that they contain younger stellar populations
than the rest of the galaxy. Moreover, several Sphs in the Virgo cluster
show central star-formation and/or disk features \citep[e.g.][]{Lisker2006,
Lisker2007}, further hinting at a possible link with BCDs and compact Irrs. 

Finally, we point out that the rotating Sphs considered here have
relatively-high luminosities ($-16 \lesssim M_{\rm R} \lesssim -18$
mag) and, thus, are at the top end of the $M - \mu_{0}$ relation
\citep[e.g.][]{Kormendy2009}. It is unclear whether Sphs with lower
luminosities and surface brightnesses, as those in the Local Group,
also show some ordered rotation \citep[see e.g.][]{Tolstoy2009}
and what their location is in the $V_{\rm R_{\rm d}}/R_{\rm d} -
\mu_{0}$ diagram. Thus, the evolutionary scenario outlined above
refers \textit{only} to the brightest Sphs found in galaxy clusters.
Typical LSB Irrs might evolve into LSB Sphs without going through
a starbursting phase \citep[e.g.][]{Weisz2011}.  

\section{Conclusions}

We studied the structure and evolution of dwarf galaxies using a
new dynamical quantity: the circular-velocity gradient $d_{R}V(0)$.
This provides a direct measure of the inner steepness of the potential
well of a galaxy and, thus, of its central dynamical mass density
(including gas, stars, and dark matter). For bulgeless, low-mass
galaxies, $d_{R}V(0)$ can be simply estimated as $V_{R_{\rm d}}
/R_{\rm d}$, where $R_{\rm d}$ is the exponential scale-length of
the stellar body. We constructed a sample of 60 low-mass galaxies
(including starbursting dwarfs, irregulars, and spheroidals),
considering objects with high-quality \hi and stellar rotation
curves from the literature. Our results can be summarized as follows.
\begin{enumerate}
 \item For gas-rich dwarfs (Irrs and BCDs), $V_{R_{\rm d}}/R_{\rm d}$
correlates with i) the central surface brightness $\mu^{i}_{0}$; ii)
the mean \hi surface density over the stellar body; and iii) the SFR
surface density.
\item Starbursting dwarfs (BCDs) are different from typical Irrs.
BCDs have high central surface brightnesses, high \hi surface densities,
and high circular-velocity gradients, implying that they have a
strong central concentration of dynamical mass (luminous and/or
dark). This suggests that the starburst is closely linked to the
inner shape of the gravitational potential and the gas density.
\item We identified several compact Irrs that have values of
$\mu^{i}_{0}$, $R_{\rm d}$, and $V_{R_{\rm d}}/R_{\rm d}$ similar
to those of BCDs. Compact Irrs are the best candidates to be the
progenitors/descendants of BCDs.
\item Rotating Sphs in the Virgo cluster follow the same correlation
between $V_{R_{\rm d}}/R_{\rm d}$ and $\mu^{i}_{0}$ determined by
Irrs and BCDs. The Sphs in our sample have values of $V_{R_{\rm d}}
/R_{\rm d}$ similar to those of BCDs and compact Irrs. This suggests
that BCDs and compact Irrs may evolve into rotating Sphs, provided
that some external mechanism removes the entire ISM.
\item Similarly to spiral galaxies, the star-formation activity in
dwarfs can be described by a law of the form $\Sigma_{\rm SFR} =
\epsilon \, \Sigma_{\rm gas}/\tau_{\rm orb}$, where $\Sigma_{\rm gas}$
considers the atomic gas only, $\tau_{\rm orb}$ is the orbital time
on the solid-body portion of the rotation curve, and the fraction
$\epsilon$ of atomic gas converted into stars during every orbit
is $\sim2\%$.

\end{enumerate}

\begin{acknowledgements}
We are grateful to Renzo Sancisi for sharing insights and ideas that
fueled this investigation. We also thank Eline Tolstoy for stimulating
discussions. FL acknowledges the Ubbo Emmius bursary program of the
University of Groningen and the Leids Kerkhoven-Bosscha Fund.
FF aknowledges financial support from PRIN MIUR 2010-2011, project
``The Chemical and Dynamical Evolution of the Milky Way and Local
Group Galaxies'', prot. 2010LY5N2T.
\end{acknowledgements}

\renewcommand\bibname{{References}}
%\addcontentsline{toc}{section}{References}
\bibliographystyle{aa}
\bibliography{bibliography5.bib}

\begin{appendix}

\section{asymmetric-drift correction\label{sec:asymDrift}}

To calculate the asymmetric-drift correction, we start from equation
(4-33) of \citet{BT87}, which describes a stationary, axisymmetric
stellar system embedded in a gravitational potential $\Phi(R, z)$:
\footnotesize
\begin{equation}\label{eq:asymDrift}
V_{\rm{circ}}^{2} = \overline{v_{\phi}}^{2} - \sigma_{\rm{R}}^{2} \bigg[ \dfrac{\partial \ln \rho}{\partial \ln R}
+ \dfrac{\partial \ln \sigma_{\rm{R}}^{2}}{\partial \ln R} + 1 - \dfrac{\sigma_{\phi}^{2}}{\sigma_{\rm{R}}^{2}}
+\dfrac{R}{\sigma_{\rm{R}}^{2}} \dfrac{\partial(\overline{v_{\rm{R}}v_{z}})}{\partial z}\bigg],
\end{equation}
\normalsize
where $v_{\rm R}$, $v_{z}$, and $v_{\phi}$ are the components
(in cylindrical coordinates) of the velocity of a star, $\rho$ is
the stellar density, $\sigma_{\rm R}^{2} = \overline{v_{\rm R}^{2}}$,
$\sigma_{\rm z}^{2} = \overline{v_{\rm z}^{2}}$, $\sigma^{2}_{\phi}
= \overline{v_{\phi}^{2}} - \overline{v_{\phi}}^{2}$, and $V_{\rm circ}
= \sqrt{R (\partial \Phi / \partial R)}$. The observed stellar rotation
curve $v_{\rm{rot}}$ provides $\overline{v_{\phi}}$. Note that
equation \ref{eq:asymDrift} does \textit{not} require that the
velocity dispersion is smaller than the rotation velocity.
Following \citet{Weijmans2008}, we write
\begin{equation}
\overline{v_{\rm{R}}v_{z}} = \kappa(\sigma_{\rm{R}}^{2} - \sigma_{z}^{2}) \dfrac{z/R}{1 - (z/R)^{2}} \qquad 0 \leq \kappa \leq 1,
\end{equation}
where $\kappa=0$ and $\kappa=1$ correspond, respectively, to the
extreme cases of a velocity ellipsoid aligned with the cilindrical
($R$, $z$, $\phi$) and spherical ($r$, $\theta$, $\phi$) coordinate
systems. Using higher-order velocity-moments of the collisionless
Boltzmann equation, \citet{Weijmans2008} obtained the following
expressions (see their Appendix A):
\begin{equation}\label{eq:sigF/sigR}
\dfrac{\sigma_{\phi}^{2}}{\sigma_{\rm{R}}^{2}} = \dfrac{1}{2} \bigg[ 1 + \alpha_{\rm{R}}
+ \kappa \dfrac{1 - \sigma_{z}^{2}/\sigma_{\rm{R}}^{2}}{1 - (z/R)^{2}} \alpha_{z}
- \dfrac{\overline{(v_{\phi} - \overline{v_{\phi}})^{3}}}{\sigma_{\rm{R}}^{2}\overline{v_{\phi}}} \bigg]
\end{equation}
and
\begin{equation}\label{eq:sigF/sigZ}
\dfrac{\sigma_{\phi}^{2}}{\sigma_{z}^{2}} = \dfrac{\kappa z^{2} (1 + \alpha_{\rm{R}})}
{\kappa z^{2}(1+\alpha_{\rm{R}}) - \alpha_{z}(R^{2}-z^{2})},
\end{equation}
where
\begin{equation}
\alpha_{\rm{R}} = \dfrac{\partial \ln \overline{v_{\phi}}}{\partial \ln R},
\quad {\rm{and}} \quad \alpha_{z} = \dfrac{\partial \ln \overline{v_{\phi}}}{\partial \ln z}.
\end{equation}
The last term of equation \ref{eq:sigF/sigR} vanishes if the velocity
ellipsoid is symmetric around $\overline{v_{\phi}}$. Since we want to
estimate $V_{\rm{circ}}$ at $R\simeq R_{\rm{d}}$, this higher-order term
can be safely ignored, as $\sigma_{\rm{R}}^{2} \simeq \sigma_{\rm{obs}}^{2}
\gg \overline{v_{\phi}}^{2}$ in the inner galaxy regions.
 
We now assume that the galaxy is in cilindrical rotation, i.e. $v_{\rm rot}
= \overline{v_{\phi}}(R)$. Observationally, it is difficult to obtain
information on the rotation velocities above the galaxy plane. A negative
velocity gradient in the vertical direction, however, would produce
an observable feature: asymmetric line profiles with a tail toward the
systemic velocity. This means that a Gauss-Hermite polynomial-fit to
the line profiles should give high values of the $h_{3}$ term. This effect
is observed, for example, in the lagging \hi haloes of spiral galaxies
\citep[e.g.][]{Fraternali2002}. The stellar, absorption-line profiles of
gas-poor dwarfs, instead, are quite symmetric and have $|h_{3}|\lesssim 0.1$
\citep{Halliday2001, Spolaor2010, Howley2013}, implying that any vertical
velocity gradient is relatively small. It is reasonable, therefore,
to assume cilindrical rotation such that $\alpha_{z}=0$.
Thus, Eq.~\ref{eq:sigF/sigZ} gives $\sigma_{\phi} = \sigma_{z}$.

Since we are interested in the inner circular-velocity gradient, we also
assume that the galaxy is in solid-body rotation, i.e. $\overline{v_{\phi}}(R)
= A \times R$. All the Sphs in our sample, indeed, show nearly solid-body
rotation curves out to the last measured point \citep[cf.][]{vanZee2004b}
and thus $\alpha_{R} = 1$. Consequently, Eq.~\ref{eq:sigF/sigR} gives
$\sigma_{\phi} = \sigma_{\rm{R}}$ (neglecting the higher-order term).
Therefore, using observationally-motivated assumptions, we find that
the Sphs in our sample can be approximated as isotropic rotators with
$\sigma_{\rm{R}} = \sigma_{z} = {\sigma_{\phi}} = \sigma_{\rm obs}$.

Finally, we assume that the scale-height of the galaxy is constant
with radius. Thus, we have $\partial\ln{\rho}/\partial\ln{R}
= \partial\ln{\Sigma}/\partial\ln{R}$, where $\Sigma$ is the surface density
profile (traced by the surface brightness profile). Assuming a Sersic profile
\citep{Sersic1963}, the asymmetric-drift-corrected circular velocity is given by
\begin{equation}\label{eq:asymFin}
V_{\rm{circ}}^{2} = v_{\rm{rot}}^{2} + \sigma_{\rm{obs}}^{2} \bigg[ \dfrac{b_{n}}{n} \bigg( \dfrac{R}{R_{\rm{eff}}} \bigg)^{1/n}
- 2 \dfrac{\partial \ln \sigma_{\rm{obs}}}{\partial \ln R} \bigg],
\end{equation}
where $R_{\rm{eff}}$ is the effective radius, $n$ is the Sersic index, and
$b_{n}$ is a constant that depends on $n$ \citep[see][]{Ciotti1991, Ciotti1999}.

For rotating Sphs, the surface brightness profile can be fitted by an
exponential-law, thus $n = 1$ and $b_{1} = 1.678$. We also assume that
$\sigma_{\rm obs}$ is constant with radius, as the observations generally
provide only the mean value $\overline{\sigma}_{\rm obs}$. Therefore,
Eq.~\ref{eq:asymFin} simplifies to $V_{\rm circ}^{2} = v_{\rm rot}^{2}
+ \overline{\sigma}_{\rm obs}^{2}(R/R_{\rm d})$, where $R_{\rm d} = 1.678
R_{\rm eff}$ is the exponential scale-length. In this case, the error
$\delta_{V/R}$ on $V_{\rm circ}/R$ is given by
\begin{equation}
\begin{split}
\delta_{V/R}^{2} &=  \bigg[ \dfrac{v_{\rm{rot}}}{V_{\rm{circ}}} \dfrac{\delta_{v_{\rm{l.o.s.}}}}{R \sin(i)} \bigg]^{2}
                   +\bigg[ \dfrac{v_{\rm{rot}}^{2}}{V_{\rm{circ}}R} \dfrac{\delta_{i}}{\tan(i)} \bigg]^{2} \\
                  &+\bigg[ \dfrac{\overline{\sigma}_{\rm{obs}}}{V_{\rm{circ}}}\dfrac{\delta_{\overline{\sigma}_{\rm{obs}}}}{R_{\rm{d}}} \bigg]^{2}
                   +\bigg[ \dfrac{V_{\rm{circ}}}{R} \dfrac{\delta_{D}}{D} \bigg]^{2},
\end{split}
\end{equation}
where $D$ and $i$ are, respectively, the galaxy distance and inclination
(cf. with Eq.~\ref{eq:errGrad}).

\section{tables}

\subsection{Tables B1 and B3 - Structural and dynamical properties of gas-rich dwarfs (BCDs and Irrs)}

\textit{Column} (1) gives the galaxy name. \\
\textit{Column} (2), (3), (4) give the assumed distance, the distance
indicator, and the corresponding reference. \\
\textit{Column} (5), (6), (7), and (8) give the $R$-band absolute magnitude
$M_{\rm R}$, the central $R$-band surface brightness $\mu^{i}_{0}$ (corrected
for inclination), the scale-length $R_{\rm d}$, and the reference for the surface
photometry. The structural parameters were derived from an exponential fit
to the outer parts of the surface brightness profiles. All the quantities
have been corrected for Galactic extinction, but not for internal extiction.\\
\textit{Column} (9) gives the galaxy inclination, derived by fitting a
tilted-ring model to the \hi velocity field and/or by building 3D model-cubes.\\
\textit{Column} (10), (11), (12), and (13) give the circular velocity
$V_{R_{\rm d}}$ at $R_{\rm{d}}$, the circular velocity $V_{\rm last}$
at the last measured point, the radius at $V_{\rm{last}}$, and the
reference for the \hi rotation curves. Values of $V_{\rm last}$ in
\textit{italics} indicate rotation curves that do not reach the flat part.\\ 
\textit{Column} (14) gives the circular-velocity gradient $V_{R_{\rm d}}/R_{\rm d}$.

\subsection{Tables B2 and B4 - Gas and star formation properties of gas-rich dwarfs (BCDs and Irrs) }

\textit{Column} (1) gives the galaxy name. \\
\textit{Column} (2) and (3) give the gas metallicity and the respective references.
Values in \textit{italics} indicate abundances derived using ``strong-line''
calibrations; we assigned to them a conservative error of 0.2 \citep[cf.][]{Berg2012}.
All the other abundances have been derived using the T$_{\rm e}$-method. \\ 
\textit{Column} (4), (5), and (6) give the \hi mass, the mean \hi surface density
within $R_{\rm opt}=3.2R_{\rm d}$ (corrected  for inclination), and the reference
for the \hi observations. \\
\textit{Column} (7), (8), and (9) give the H$\alpha$+\NII equivalent width, the
H$\alpha$ star formation rate (SFR), and the reference for the H$\alpha$ observations.
SFRs have been calculated using the \citet{Kennicutt1998} calibration and have not
been corrected for internal extinction. \\
\textit{Column} (10) and (11) give the ratio SFR/$M_{\rm bar}$ and the SFR surface
density ($\Sigma_{\rm{SFR}} = \rm{SFR}/\pi R_{\rm{opt}}^{2}$). $M_{\rm bar}$
has been estimated using the baryonic Tully-Fisher relation as calibrated by
\citet{McGaugh2012} with an accuracy of $\sim$10$\%$. Values of SFR/$M_{\rm bar}$
in \textit{italics} indicate galaxies with rotation curves that do not reach
the flat part, thus they may be slightly underestimated. \\
\textit{Column} (12) and (13) give the global and local gas depletion times
$\tau_{\rm global}$ and $\tau_{\rm local}$. $\tau_{\rm global}$ considers
the total atomic gas mass of the galaxy, whereas $\tau_{\rm local}$ considers
the atomic gas mass within $R_{\rm opt}$.

\subsection{Table B5 - Structural and dynamical properties of gas-poor dwarfs (Sphs)}

\textit{Column} (1) gives the galaxy name. \\
\textit{Column} (2), (3), (4), (5), and (6) give the $R$-band absolute magnitude,
the inclination $i$, the central surface brightness (corrected for $i$), the
scale-length, and the reference for the surface photometry. The inclination was
estimated from the observed ellipticities using Eq.~\ref{eq:incl} with $q_{0}=0.35$.
The structural parameters were derived from an exponential fit to the outer parts
of the surface brightness profiles. All the quantities have been corrected for
Galactic extinction, but not for internal extinction.\\
\textit{Column} (7), (8), (9), (10), and (11) give the rotation velocity at
$R_{\rm d}$, the rotation velocity $v_{\rm last}$ at the last measured point,
the radius at $v_{\rm last}$, the mean velocity dispersion, and the reference
for the stellar spectroscopy. All rotation velocities have been corrected for
inclination.\\
\textit{Column} (12), (13) and (14) give the circular velocity at $R_{\rm d}$,
the circular velocity at $R_{\rm last}$, and the circular-velocity gradient.
All the circular velocities have been corrected for asymmetric-drift (see
Appendix \ref{sec:asymDrift}).

\begin{table*}[t]
\caption{Sample of starbursting dwarfs. Structural and kinematic properties.}
\centering
\resizebox{18.5cm}{!}{
\begin{tabular}{l | c c c | c c c c | c c c c c | c}
\hline
\hline
Name & Dist & Method & Ref. & $M_{\rm{R}}$ & $\mu^{i}_{0, R}$ & $R_{\rm{d}}$ & Ref. & $i$ & $V_{R_{\rm{d}}}$ & $V_{\rm{last}}$ & $R_{\rm{last}}$
& Ref. & $V_{R_{\rm{d}}}/R_{\rm{d}}$ \\
     & (Mpc)   &        &      & (mag)          & (mag/$''^{2}$) & (kpc)     &      & ($^{\circ}$) & (km/s) & (km/s)            & (kpc)              
&      & (km/s/kpc)           \\
(1)                & (2)          & (3)  &(4)& (5)           & (6)  & (7)  &(8)& (9)      & (10)      & (11)     & (12)  & (13) &(14)     \\ 
\hline
NGC 1705\tablefootmark{\alpha}
                   & 5.1$\pm$0.6  & TRGB & a & -16.3$\pm$0.2 & 21.1$\pm$0.1 & 0.46 & j & 45$\pm$5  & 58$\pm$8  & 71$\pm$10& 5.7    & j & 126$\pm$25 \\
NGC 2366           & 3.2$\pm$0.4  & Ceph & b & -16.6$\pm$0.3 & 22.5$\pm$0.2 & 1.36 & k & 68$\pm$5  & 30$\pm$2  & 49$\pm$6 & 5.9    & j & 22$\pm$3 \\
NGC 2915\tablefootmark{\beta}
                   & 3.8$\pm$0.4  & TRGB & c & -15.9$\pm$0.2 & 20.5$\pm$0.1 & 0.46 & l & 52$\pm$3  & 32$\pm$15 & 85$\pm$5 & 9.3    & n & 70$\pm$33 \\ 
NGC 4068           & 4.3$\pm$0.1  & TRGB & d & -15.7$\pm$0.1 & 21.8$\pm$0.1 & 0.56 & k & 44$\pm$6  & 13$\pm$4  &\textit{39$\pm$5}& 2.3 & j & 23$\pm$8 \\ 
NGC 4214\tablefootmark{\alpha}
                   & 2.7$\pm$0.2  & TRGB & d & -17.8$\pm$0.2 & 20.3$\pm$0.1 & 0.70 & k & 30$\pm$10 & 64$\pm$8  & 81$\pm$9 & 4.8    & j & 91$\pm$31 \\
NGC 6789           & 3.6$\pm$0.2  & TRGB & e & -15.1$\pm$0.1 & 21.1$\pm$0.2 & 0.23 & j & 43$\pm$7  & 23$\pm$4  &\textit{57$\pm$9} & 0.7 & j & 100$\pm$22 \\
UGC 4483           & 3.2$\pm$0.2  & TRGB & f & -13.0$\pm$0.1 & 22.7$\pm$0.1 & 0.20 & j & 58$\pm$3  & 10$\pm$3  & 19$\pm$2 & 1.2    & o & 50$\pm$15 \\
I Zw 18\tablefootmark{\gamma}
                   & 18.2$\pm$1.4 & TRGB & g & -15.0$\pm$0.2 & 20.5$\pm$0.2 & 0.15 & m & 70$\pm$4  & 25$\pm$10 & 38$\pm$4 & 0.8    & p & 167$\pm$60 \\
I Zw 36            & 5.9$\pm$0.5  & TRGB & h & -14.9$\pm$0.1 & 22.5$\pm$0.2 & 0.29 & j & 67$\pm$3  & 16$\pm$2  & 29$\pm$2 & 1.1    & j & 55$\pm$8 \\
%SBS~1415+347       & 13.5$\pm$1.4 & TRGB & i & -15.9$\pm$0.2 & 22.6$\pm$0.2 & 0.74 & j & 66$\pm$3  & 20$\pm$4  & 18$\pm$2 & 3.6    & j & 27$\pm$7 \\
\hline
\hline
\end{tabular}
}
\tablefoot{\tablefoottext{\alpha}{The \hi disk is warped, the inclination is given for $R\simeq R_{\rm{d}}$.}\\
\tablefoottext{\beta}{We adopted the rotation curve derived by \citet{Elson2010} assuming a constant inclination of $52^{\circ}$}.\\
\tablefoottext{\gamma}{The optical parameters have been derived by \citet{Papaderos2002} after subtracting the H$\alpha$ emission.}}
\tablebib{(a)~\citet{Tosi2001}; (b)~\citet{Tolstoy1995}; (c)~\citet{Karachentsev2003}; (d)~\citet{Jacobs2009};
(e)~\citet{Drozdovsky2001}; (f)~\citet{Izotov2002}; (g)~\citet{Aloisi2007}; (h)~\citet{Schulte2001}; (i)~\citet{Aloisi2005};
(j)~Lelli et al. (submitted); (k)~\citet{Swaters2002b}; (l)~\citet{Meurer1994}; (m)~\citet{Papaderos2002}; (n)~\citet{Elson2010};
(o)~\citet{Lelli2012b}; (p)~\citet{Lelli2012a}.}
\label{tab:BCDs}
\end{table*}

\begin{table*}[t]
\caption{Sample of starbursting dwarfs. Gas and star-formation properties.}
\centering
\resizebox{18.5cm}{!}{
\begin{tabular}{l | c c | c c c | c c c | c c c c}
\hline
\hline
Name  & log(O/H) & Ref.& $M_{\hi}$         & $\avghi$             & Ref. & EW(H$\alpha$)        & SFR            & Ref.& SFR/$M_{\rm bar}$
& log($\Sigma_{\rm{SFR}}$)      & $\tau_{\rm{global}}$ & $\tau_{\rm{local}}$ \\
      & +12            &     & (10$^{8}M_{\odot}$) & ($M_{\odot}$/pc$^{2}$) &      & ($\mathring{\rm A}$) & ($M_{\odot}$/yr) &     & (Gyr$^{-1}$) 
& ($M_{\odot}$/yr/kpc$^{2}$) & (Gyr)        & (Gyr) \\
(1)          & (2)          & (3)           &(4)& (5)           & (6)         &(7)& (8)        & (9)  &(10)& (11) & (12) &(13)\\ 
\hline
NGC 1705     & 8.21$\pm$0.05 & a & 1.1$\pm$0.3   & 7.0$\pm$0.9 & d & 109$\pm$7  & 0.075 & f & 0.063 & -1.96 & 1.9 & 0.8 \\
NGC 2366     & 7.91$\pm$0.05 & b & 6.2$\pm$1.7   & 6.0$\pm$1.4 & d & 149$\pm$38 & 0.100 & f & 0.369 & -2.77 & 8.2 & 4.7 \\
NGC 2915     & 8.27$\pm$0.20 & a & 4.4$\pm$1.0   & 5.8$\pm$0.7 & e &  38$\pm$5  & 0.028 & f & 0.011 & -2.39 & 20.9& 1.9 \\
NGC 4068     & ...           &...& 1.5$\pm$0.2   & 8.1$\pm$1.1 & d &  28$\pm$5  & 0.014 & f & \textit{0.129} & -2.89 & 14.2 & 8.3 \\
NGC 4214     & 8.22$\pm$0.05 & a & 4.3$\pm$0.8   & 6.9$\pm$1.0 & d &  62$\pm$7  & 0.104 & f & 0.051 & -2.18 & 5.5 & 1.4 \\
NGC 6789     & ...           &...& 0.18$\pm$0.03 & 7.3$\pm$1.1 & d &  23$\pm$3  & 0.004 & f & \textit{0.008} & -2.63 & 6.0 & 4.1 \\
UGC 4483     & 7.56$\pm$0.03 & b & 0.29$\pm$0.05 & 7.9$\pm$1.0 & d & 130$\pm$20 & 0.003 & f & 0.490 & -2.63 & 12.9 & 4.5 \\
I Zw 18      & 7.18$\pm$0.01 & c & 2.1$\pm$0.4   &22.0$\pm$4.8 & d & 679$\pm$68 & 0.111 & g & 1.133 & -0.81 & 2.5 & 0.2 \\
I Zw 36      & 7.77$\pm$0.01 & c & 0.7$\pm$0.1   & 5.5$\pm$0.9 & d & 335$\pm$17 & 0.041 & f & 1.233 & -1.82 & 2.3 & 0.5 \\
%SBS~1415+347 & 7.59$\pm$0.01 & c & 2.1$\pm$0.5   & 3.8$\pm$0.2 & d & 207$\pm$25 & 0.087 & f & 17.63 & -2.31 & 3.2 & 1.0 \\
\hline
\hline
\end{tabular}
}
\tablebib{(a)~\citet{Berg2012}; (b)~\citet{Croxall2009}; (c)~\citet{Izotov1999}; (d)~Lelli et al. (submitted);
(e)~\citet{Elson2010}; (f)~\citet{Kennicutt2008}; (g)~\citet{GilDePaz2003}.}
\label{tab:BCDs2}
\end{table*}

\begin{table*}[htb!]
\caption{Sample of irregulars. Structural and kinematic properties.}
\centering
\resizebox{18.5cm}{!}{
\begin{tabular}{l | c c c | c c c c | c c c c c | c}
\hline
\hline
Name & Dist & Method & Ref. & $M_{\rm{R}}$ & $\mu^{i}_{0, R}$ & $R_{\rm{d}}$ & Ref. & $i$ & $V_{R_{\rm{d}}}$ & $V_{\rm{last}}$ & $R_{\rm{last}}$
& Ref. & $V_{R_{\rm{d}}}/R_{\rm{d}}$ \\
     & (Mpc)   &        &      & (mag)          & (mag/$''^{2}$) & (kpc)     &      & ($^{\circ}$) & (km/s) & (km/s)            & (kpc)              
&      & (km/s/kpc)           \\
(1)                & (2)          & (3)  &(4)& (5)           & (6)  & (7)  &(8)& (9)      & (10)      & (11)     & (12)  & (13) & (14)     \\ 
\hline
UGC 731             & 11.8$\pm$4.3 & TF   & a & -17.1$\pm$0.8 & 22.9$\pm$0.1 & 2.62 & h & 57$\pm$3  & 50$\pm$4  & 74$\pm$4 & 10.3 & k & 19$\pm$7 \\
UGC 2455\tablefootmark{\alpha}
                    & 6.4$\pm$1.2  & TF   & a & -17.7$\pm$0.4 & 20.1$\pm$0.1 & 0.72 & h & 51$\pm$3  & 23$\pm$10 & \textit{61$\pm$4} & 3.7 & k & 32$\pm$15 \\
UGC 3371            & 21.9$\pm$4.9 & TF   & b & -18.6$\pm$0.4 & 23.2$\pm$0.1 & 5.58 & h & 49$\pm$3  & 51$\pm$3  & 86$\pm$3 & 17.4 & k & 9$\pm$2 \\
UGC 3711            & 8.2$\pm$1.5  & TF   & a & -17.4$\pm$0.4 & 21.1$\pm$0.1 & 0.87 & h & 60$\pm$3  & 79$\pm$4  & 95$\pm$4 & 3.6  & k & 91$\pm$17 \\
UGC 3817            & 8.3$\pm$3.1  & TF   & a & -14.8$\pm$0.8 & 22.7$\pm$0.1 & 0.64 & h & 30$\pm$3  & 16$\pm$5  & \textit{45$\pm$5} & 2.4 & k & 25$\pm$12\\
UGC 3966            & 7.4$\pm$1.4  & TF   & a & -15.2$\pm$0.4 & 22.5$\pm$0.1 & 0.67 & h & 41$\pm$3  & 38$\pm$5  & 50$\pm$5 & 2.7   & k & 57$\pm$13 \\
UGC 4173            & 16.7$\pm$3.1 & TF   & a & -17.7$\pm$0.4 & 24.1$\pm$0.1 & 4.97 & h & 40$\pm$3  & 36$\pm$5	& \textit{57$\pm$5} & 12.2 & k & 7$\pm$2 \\
UGC 4325            & 10.0$\pm$1.8 & TF   & a & -18.0$\pm$0.4 & 21.4$\pm$0.1 & 1.74 & h & 41$\pm$3  & 73$\pm$3  & 92$\pm$3 & 5.8   & k & 42$\pm$8 \\
UGC 4499            & 12.8$\pm$2.4 & TF   & a & -17.7$\pm$0.4 & 21.8$\pm$0.1 & 1.33 & h & 50$\pm$3  & 38$\pm$4  & 74$\pm$3 & 8.4   & k & 29$\pm$6 \\
UGC 4543            & 30.0$\pm$5.5 & TF   & a & -19.2$\pm$0.4 & 22.4$\pm$0.1 & 3.34 & h & 46$\pm$3  & 59$\pm$4  & 67$\pm$4 & 17.4  & k & 18$\pm$4 \\
UGC 5272            & 6.5$\pm$1.2  & TF   & a & -15.2$\pm$0.4 & 22.0$\pm$0.1 & 0.68 & h & 59$\pm$3  & 19$\pm$4  & \textit{45$\pm$4} & 1.9 & k & 28$\pm$8 \\
UGC 5414            & 9.4$\pm$1.7  & TF   & a & -17.4$\pm$0.4 & 21.8$\pm$0.1 & 1.38 & h & 55$\pm$3  & 35$\pm$3  & \textit{61$\pm$2} & 4.1 & k & 25$\pm$5 \\
UGC 5721            & 5.9$\pm$1.1  & TF   & a & -16.3$\pm$0.4 & 20.3$\pm$0.1 & 0.39 & h & 61$\pm$3  & 39$\pm$3  & 79$\pm$3 & 6.4   & k & 100$\pm$20 \\
UGC 5829            & 8.0$\pm$1.5  & TF   & a & -17.0$\pm$0.4 & 22.6$\pm$0.1 & 1.50 & h & 34$\pm$3  & 34$\pm$5	& \textit{69$\pm$5} & 6.4 & k & 23$\pm$6 \\
UGC 5846            & 19.3$\pm$3.6 & TF   & a & -16.9$\pm$0.4 & 22.8$\pm$0.1 & 1.77 & h & 30$\pm$3  & 30$\pm$6	& 51$\pm$6 & 5.6   & k & 17$\pm$5 \\
UGC 5918            & 7.1$\pm$1.3  & TF   & a & -15.2$\pm$0.4 & 23.8$\pm$0.1 & 1.59 & h & 46$\pm$3  & 30$\pm$4 	& \textit{45$\pm$4} & 4.1   & k & 19$\pm$4 \\
UGC 6399            & 18.4$\pm$3.0 & TF   & b & -18.0$\pm$0.3 & 22.3$\pm$0.2 & 2.14 & i & 75$\pm$2  & 53$\pm$7  & 88$\pm$5 & 8.0   & l & 26$\pm$5 \\
UGC 6446            & 18.0$\pm$3.0 & TF   & b & -18.5$\pm$0.4 & 22.2$\pm$0.1 & 2.70 & i & 51$\pm$3  & 62$\pm$4  & 82$\pm$4 & 15.7  & l & 23$\pm$4 \\
UGC 6955            & 16.1$\pm$2.7 & TF   & b & -18.1$\pm$0.4 & 23.0$\pm$0.1 & 3.91 & h & 64$\pm$2  & 49$\pm$3  & 82$\pm$2 & 15.6  & m & 12$\pm$2 \\ 
UGC 7232            & 2.8$\pm$0.5  & TF   & a & -14.8$\pm$0.4 & 20.7$\pm$0.1 & 0.20 & h & 59$\pm$3  & 19$\pm$4  & \textit{44$\pm$4} & 0.8 & k & 95$\pm$26 \\ 
UGC 7261            & 7.9$\pm$1.5  & TF   & a & -17.3$\pm$0.4 & 22.0$\pm$0.1 & 1.34 & h & 30$\pm$3  & 66$\pm$6  & 76$\pm$6 & 4.0   & k & 49$\pm$11 \\
UGC 7323            & 5.8$\pm$1.0  & TF   & b & -18.2$\pm$0.4 & 21.3$\pm$0.1 & 1.51 & h & 47$\pm$3  & 49$\pm$4  & \textit{86$\pm$4} & 4.2   & k & 32$\pm$6 \\
UGC 7524\tablefootmark{\beta}
                    & 4.3$\pm$0.4  & Ceph & d & -18.6$\pm$0.2 & 22.1$\pm$0.1 & 2.82 & h & 46$\pm$3  & 58$\pm$4 	& 79$\pm$4 & 9.7   & k & 21$\pm$3 \\
UGC 7559            & 5.0$\pm$0.2  & TRGB & e & -14.6$\pm$0.1 & 23.9$\pm$0.1 & 1.08 & h & 61$\pm$3  & 21$\pm$4 	& 33$\pm$4 & 3.3   & k & 19$\pm$4 \\
UGC 7577            & 2.6$\pm$0.1  & TRGB & e & -15.0$\pm$0.1 & 22.4$\pm$0.1 & 0.64 & h & 63$\pm$3  & 8$\pm$3   & \textit{18$\pm$3} & 1.7 & k & 12$\pm$5 \\
UGC 7603            & 10.5$\pm$1.7 & TF   & b & -17.8$\pm$0.3 & 21.3$\pm$0.3 & 1.06 & h & 78$\pm$3  & 30$\pm$3  & 64$\pm$3 & 9.2   & k & 28$\pm$5\\
UGC 7690\tablefootmark{\gamma}
                    & 7.5$\pm$1.4  & TF   & a & -16.9$\pm$0.4 & 20.2$\pm$0.1 & 0.43 & h & 41$\pm$3  & 44$\pm$4  & 56$\pm$4 & 3.8   & k & 102$\pm$22\\
UGC 7866            & 4.6$\pm$0.2  & TRGB & e & -15.1$\pm$0.1 & 22.1$\pm$0.1 & 0.55 & h & 44$\pm$3  & 17$\pm$4  & \textit{33$\pm$4} & 2.3 & k & 31$\pm$8 \\
UGC 7916            & 7.2$\pm$2.6  & TF   & a & -14.6$\pm$0.8 & 24.4$\pm$0.2 & 1.47 & h & 74$\pm$3  & 21$\pm$3  & \textit{36$\pm$3} & 3.7 & k & 14$\pm$5 \\
UGC 7971            & 8.0$\pm$1.5  & TF   & a & -17.0$\pm$0.4 & 21.3$\pm$0.1 & 0.89 & h & 38$\pm$3  & 22$\pm$5 	& \textit{45$\pm$5} & 2.9 & k & 25$\pm$7 \\
UGC 8320            & 4.6$\pm$0.2  & TRGB & e & -15.9$\pm$0.1 & 22.3$\pm$0.1 & 0.79 & h & 61$\pm$3  & 21$\pm$4  & 49$\pm$4 & 4.0   & m & 27$\pm$5 \\
UGC 8490\tablefootmark{\gamma} 
                    & 4.6$\pm$0.6  & TRGB & f & -17.2$\pm$0.3 & 20.4$\pm$0.1 & 0.65 & h & 50$\pm$3  & 48$\pm$4  & 78$\pm$4 & 10.0  & k & 74$\pm$12 \\
UGC 8837            & 7.2$\pm$0.1  & TRGB & e & -16.5$\pm$0.1 & 23.7$\pm$0.3 & 1.73 & h & 80$\pm$3  & 25$\pm$3  & \textit{48$\pm$4} & 4.2 & k & 14$\pm$2 \\
UGC 9211            & 14.7$\pm$2.7 & TF   & a & -16.5$\pm$0.4 & 22.7$\pm$0.1 & 1.38 & h & 44$\pm$3  & 35$\pm$4	& 65$\pm$4 & 9.6   & k & 25$\pm$6 \\
UGC 9992            & 11.2$\pm$2.1 & TF   & a & -16.0$\pm$0.4 & 21.9$\pm$0.1 & 0.86 & h & 30$\pm$3  & 28$\pm$6  & 34$\pm$6 & 4.1   & k & 33$\pm$10 \\
UGC 10310           & 15.8$\pm$2.9 & TF   & a & -17.9$\pm$0.4 & 21.9$\pm$0.1 & 1.91 & h & 34$\pm$3  & 44$\pm$5  & 74$\pm$5 & 9.2   & k & 23$\pm$5 \\
UGC 11557           & 23.7$\pm$4.4 & TF   & a & -19.0$\pm$0.4 & 20.9$\pm$0.1 & 3.01 & h & 30$\pm$3  & 53$\pm$6 	& 85$\pm$6 & 10.3  & k & 18$\pm$4 \\
UGC 11707           & 15.7$\pm$3.0 & TF   & a & -18.2$\pm$0.4 & 23.1$\pm$0.2 & 4.41 & h & 68$\pm$3  & 68$\pm$7  & 100$\pm$3 & 15.0 & k & 15$\pm$3 \\
UGC 12060           & 15.1$\pm$2.8 & TF   & a & -17.7$\pm$0.4 & 21.8$\pm$0.1 & 1.52 & h & 40$\pm$3  & 61$\pm$4  & 74$\pm$4 & 9.9   & k & 40$\pm$8 \\
UGC 12632           & 9.2$\pm$1.7  & TF   & a & -17.5$\pm$0.4 & 23.3$\pm$0.1 & 3.78 & h & 46$\pm$3  & 58$\pm$4  & 76$\pm$3 & 11.4  & k & 15$\pm$3 \\
UGC 12732           & 12.4$\pm$2.3 & TF   & a & -17.8$\pm$0.4 & 22.4$\pm$0.1 & 2.13 & h & 39$\pm$3  & 53$\pm$5  & \textit{98$\pm$5} & 14.4  & k & 25$\pm$5 \\
WLM\tablefootmark{\delta}
                    & 1.0$\pm$0.1  & Ceph & g & -14.1$\pm$0.2 & 22.5$\pm$0.2 & 0.57 & j & 70$\pm$4  & 19$\pm$2  & \textit{43$\pm$4} & 3.7   & n & 33$\pm$5 \\
NGC 6822\tablefootmark{\delta}
                    & 0.5$\pm$0.1  & Ceph & g & -15.0$\pm$0.4 & 21.3$\pm$0.1 & 0.57 & j & 59$\pm$3 & 20$\pm$1   & 56$\pm$3 & 4.7   & o & 35$\pm$7 \\
\hline
\hline
\end{tabular}
}
\tablefoot{\footnotesize
\tablefoottext{\alpha}{the \hi line profiles are very broad and complex; we assigned a larger error to $V_{R_{\rm{d}}}$ than given by \citet{Swaters2009}.}\\
\tablefoottext{\beta}{the galaxy is kinematically lopsided, but the \hi kinematics is quite symmetric in the inner regions.}\\
\tablefoottext{\gamma}{the \hi disk is warped, the inclination is given for $R\simeq R_{\rm{d}}$.}\\
\tablefoottext{\delta}{the $V$-band values from \citet{Hunter2006} have been converted to $R$-band assuming $V-R=0.5$.} }
\tablebib{(a)~\citet{Tully1988}; (b)~\citet{Tully2009}; (c)~\citet{Hoessel1998}; (d)~\citet{Thim2004};
(e)~\citet{Jacobs2009}; (f)~\citet{Drozdovsky2000}; (g)~\citet{Bono2010}; (h)~\citet{Swaters2002b}; (i)~\citet{Tully1996};
(j)~\citet{Hunter2006}; (k)~\citet{Swaters2009}; (l)~\citet{Verheijen2001}; (m)~\citet{Broeils1992}; (n)~\citet{Kepley2007};
(o)~\citet{Weldrake2003}.}
\label{tab:Irrs}
\end{table*}

\begin{table*}[htb!]
\caption{Sample of irregulars. Gas and star-formation properties.}
\centering
\resizebox{18.5cm}{!}{
\begin{tabular}{l | c c | c c c | c c c | c c c c}
\hline
\hline
Name  & log(O/H)  & Ref.& $M_{\hi}$         & $\avghi$             & Ref. & EW(H$\alpha$)  & SFR            & Ref.& SFR/$M_{\rm bar}$
& log($\Sigma_{\rm{SFR}}$)      & $\tau_{\rm{global}}$ & $\tau_{\rm{local}}$ \\
      & +12       &     & (10$^{8}M_{\odot}$) & ($M_{\odot}$/pc$^{2}$) &      & ($\mathring{\rm A}$) & ($M_{\odot}$/yr) &     & (Gyr$^{-1}$) 
& ($M_{\odot}$/yr/kpc$^{2}$) & (Gyr)        & (Gyr) \\
(1)   & (2)          & (3)           &(4)& (5)         & (6)        &(7)& (8)        & (9)  &(10)&(11)& (12) &(13)\\ 
\hline
UGC 731   & \textit{8.46$\pm$0.20} & a & 16.1$\pm$11.8 & 5.5$\pm$0.7  & g & ...      & ...   &...&...&...& ...   &...\\
UGC 2455  & \textit{8.39$\pm$0.20} & a & 6.9$\pm$2.7   & 10.1$\pm$1.2 & g & 49$\pm$4 & 0.161 & l & \textit{0.247} & -2.01 &  5.7 & 1.4\\
UGC 3371  & \textit{8.48$\pm$0.20} & a & 35.6$\pm$16.3 & 3.2$\pm$0.4  & g & 24$\pm$2 & 0.041 & m & 0.016 & -4.39 & 115.5 &104.0\\
UGC 3711  & ...                    &...& 6.2$\pm$2.4   & 7.7$\pm$1.0  & g & 44$\pm$4 & 0.097 & l & 0.025 & -2.40 &  8.5 & 2.6 \\
UGC 3817  & ...                    &...& 2.1$\pm$1.6   & 4.5$\pm$0.5  & g & 59$\pm$8 & 0.008 & l & \textit{0.041} & -3.22 & 34.9 & 9.9 \\
UGC 3966  & \textit{8.15$\pm$0.20} & a & 3.2$\pm$1.2   & 6.4$\pm$0.7  & g & 20$\pm$4 & 0.005 & l & 0.017 & -3.46 & 85.1 & 24.6 \\
UGC 4173  & ...                    &...& 23.8$\pm$8.0  & 3.0$\pm$0.3  & g & 60$\pm$6 & 0.056 & m & \textit{0.113} & -4.15 & 56.5 & 56.5 \\
UGC 4325  & 8.15$\pm$0.05          & b & 7.4$\pm$2.7   & 6.6$\pm$0.7  & g & 32$\pm$4 & 0.075 & l & 0.022 & -3.11 & 12.9 & 11.4 \\
UGC 4499  & ...                    &...& 11.5$\pm$4.5  & 7.2$\pm$0.8  & g & 49$\pm$5 & 0.082 & m & 0.058 & -2.84 & 18.6 & 6.6 \\
UGC 4543  & ...                    &...& 72.2$\pm$27.4 & 5.4$\pm$0.6  & g & ...      & ...   &...&...&...& ...  &...\\
UGC 5272  & 7.87$\pm$0.05          & b & 1.9$\pm$0.7   & 8.9$\pm$1.2  & g & 45$\pm$4 & 0.016 & l & \textit{0.083} & -2.97 & 15.8 & 11.0 \\
UGC 5414  & ...                    &...& 5.7$\pm$2.1   & 6.0$\pm$0.7  & g & 49$\pm$5 & 0.051 & m & \textit{0.078} & -3.08 & 14.9 & 9.6 \\
UGC 5721  & \textit{8.32$\pm$0.20} & a & 5.1$\pm$2.0   & 11.7$\pm$1.6 & g & 53$\pm$4 & 0.037 & l & 0.020 & -2.12 & 18.3 & 2.1 \\
UGC 5829  & \textit{8.30$\pm$0.10} & c & 8.8$\pm$3.4   & 6.7$\pm$0.7  & g & 38$\pm$4 & 0.049 & l & \textit{0.046} & -3.17 & 23.9 & 13.2 \\
UGC 5846  & ...                    &...& 15.3$\pm$5.9  & 5.1$\pm$0.5  & g & ...      & ...   &...&...&...& ...   &...\\
UGC 5918  & 7.84$\pm$0.04          & d & 2.5$\pm$0.9   & 2.6$\pm$0.3  & g & 19$\pm$5 & 0.003 & l & 0.016 & -4.43 & 110.8 & 93.7\\
UGC 6399  & ...                    &...& 8.4$\pm$2.7   & ...          & h & 25$\pm$2 & 0.067 & m & 0.024 & -3.34 & 16.7 & ...\\
UGC 6446  & ...                    &...& 31.0$\pm$10.3 & 5.5$\pm$0.6  & h & 80$\pm$8 & 0.117 & m & 0.055 & -3.30 & 35.2 & 14.7 \\
UGC 6955  & \textit{8.26$\pm$0.20} & a & 20.8$\pm$7.3  & ...          & i & 20$\pm$5 & 0.047 & m & 0.022 & -4.02 & 58.9 & ... \\
UGC 7232  & ...                    &...& 0.4$\pm$0.2   & 7.8$\pm$1.0  & g & 22$\pm$5 & 0.003 & l & \textit{0.017} & -2.63 & 17.7 & 4.4 \\
UGC 7261  & ...                    &...& 5.0$\pm$2.0   & 5.2$\pm$0.5  & g & 42$\pm$4 & 0.075 & l & 0.048 & -2.89 &  8.9 & 5.3 \\
UGC 7323  & ...                    &...& 3.8$\pm$1.4   & 4.1$\pm$0.5  & g & 18$\pm$3 & 0.038 & l & 0.015 & -3.29 & 13.3 & 10.5\\
UGC 7524  & 8.09$\pm$0.15          & e & 14.6$\pm$3.0  & 3.9$\pm$0.4  & g & 27$\pm$4 & 0.079 & l & 0.043 & -3.51 & 24.6 & 16.8\\
UGC 7559  & ...                    &...& 1.8$\pm$0.2   & 3.6$\pm$0.5  & g & 38$\pm$5 & 0.008 & l & 0.143 & -3.67 & 29.9 & 22.5\\
UGC 7577  & 7.97$\pm$0.06          & b & 0.45$\pm$0.06 & 2.1$\pm$0.3  & g &  9$\pm$2 & 0.002 & l & \textit{0.405} & -3.82 & 29.9 & 18.4\\
UGC 7603  & ...                    &...& 12.8$\pm$4.3  & 6.3$\pm$1.7  & g & 35$\pm$4 & 0.081 & l & 0.103 & -2.65 & 21.0 & 3.7\\
UGC 7690  & ...                    &...& 3.3$\pm$1.3   & 8.8$\pm$1.0  & g & 24$\pm$4 & 0.024 & l & 0.052 & -2.39 & 18.3 & 2.9 \\
UGC 7866  & ...                    &...& 1.2$\pm$0.2   & 4.7$\pm$0.5  & g & 49$\pm$4 & 0.012 & l & \textit{0.215} & -2.91 & 13.3 & 5.1\\
UGC 7916  & ...                    &...& 2.6$\pm$1.9   & 3.1$\pm$0.6  & g & 63$\pm$22& 0.014 & l & \textit{0.177} & -3.70 & 24.7 & 20.5\\
UGC 7971  & \textit{8.43$\pm$0.20} & a & 2.5$\pm$0.9   & 5.7$\pm$0.6  & g & 25$\pm$4 & 0.017 & l & \textit{0.088} & -3.18 & 19.6 & 11.4\\
UGC 8320  & \textit{8.29$\pm$0.20} & a & 3.5$\pm$0.5   & ...          & i & 22$\pm$4 & 0.008 & l & 0.029 & -3.40 & 58.2 & ... \\
UGC 8490  & ...                    &...& 7.0$\pm$2.0   & 9.1$\pm$1.1  & g & 54$\pm$5 & 0.063 & l & 0.036 & -2.33 & 14.8 & 2.6 \\
UGC 8837  & 7.87$\pm$0.07          & b & 3.2$\pm$0.3   & 1.8$\pm$0.6  & g & 37$\pm$4 & 0.022 & l & \textit{0.088} & -3.64 & 19.3 &10.5 \\
UGC 9211  & ...                    &...& 14.2$\pm$5.4  & 6.2$\pm$0.7  & g & 57$\pm$12& 0.036 & l & 0.043 & -3.23 & 52.5 & 14.0 \\
UGC 9992  & 7.88$\pm$0.12          & c & 3.5$\pm$1.4   & 5.3$\pm$0.5  & g & 20$\pm$4 & 0.009 & l & 0.143 & -3.42 & 51.7 & 18.6 \\
UGC 10310 & \textit{8.31$\pm$0.20} & a & 12.9$\pm$4.9  & 6.2$\pm$0.7  & g & 63$\pm$6 & 0.147 & m & 0.104 & -2.90 & 11.7 & 6.6\\
UGC 11557 & ...                    &...& 25.0$\pm$9.6  & 5.0$\pm$0.5  & g & 35$\pm$3 & 0.486 & m & 0.198 & -2.78 &  6.8 & 4.0\\
UGC 11707 & ...                    &...& 36.3$\pm$14.3 & 5.2$\pm$0.8  & g & ...      & ...   &...&...&...& ...  &...\\
UGC 12060 & \textit{8.34$\pm$0.20} & a & 16.7$\pm$6.4  & 4.0$\pm$0.4  & g & ...      & ...   &...&...&...& ...  &...\\
UGC 12632 & \textit{8.34$\pm$0.20} & a & 15.4$\pm$5.9  & 3.4$\pm$0.4  & g & 40$\pm$6 & 0.060 & l & 0.038 & -3.88 & 34.1 &34.6\\
UGC 12732 & ...                    &...& 32.3$\pm$12.4 & 4.7$\pm$0.5  & g & 88$\pm$9 & 0.086 & m & \textit{0.020} & -3.23 & 49.9 &10.6\\
WLM       & 7.83$\pm$0.06          & b & 0.7$\pm$0.1   & ...          & j & 25$\pm$9 & 0.002 & l & \textit{0.012} & -3.72 & 46.5 & ...\\
NGC 6822\tablefootmark{\alpha}
          & 8.11$\pm$0.10          & f & 1.5$\pm$0.6   & ...          & k & 47$\pm$12& 0.011 & l & 0.024 & -2.98 & 18.1 & ...\\
\hline
\hline
\end{tabular}
}
\tablefoot{\tablefoottext{\alpha}{\citet{Weldrake2003} do not provide the total \hi flux, thus we used
the single-dish value from \citet{Koribalski2004} to calculate $M_{\hi}$.}}
\tablebib{(a)~\citet{Hunter1999}; (b)~\citet{Berg2012}; (c)~\citet{vanZee2006}; (d)~\citet{Croxall2009};
(e)~\citet{Esteban2009}; (f)~\citet{Lee2005}; (g)~\citet{Swaters2002a}; (h)~\citet{Verheijen2001};
(i)~\citet{Broeils1992}; (j)~\citet{Kepley2007}; (k)~\citet{Weldrake2003}; (l)~\citet{Kennicutt2008};
(m)~\citet{James2004}.}
\label{tab:Irrs2}
\end{table*}

\begin{table*}[t]
\caption{Sample of rotating spheroidals in the Virgo cluster.}
\centering
\resizebox{18.5cm}{!}{
\begin{tabular}{l | c c c c c | c c c c c | c c c}
\hline
\hline
Name & $M_{\rm{R}}$ & $i$        & $\mu^{i}_{0, R}$ & $R_{\rm{d}}$ & Ref.& $v_{R_{\rm{d}}}$ & $v_{\rm{last}}$& $R_{\rm{last}}$ &
$\overline{\sigma}_{\rm{obs}}$ & Ref. & $V_{R_{\rm{d}}}$ & $V_{\rm{last}}$ & $V_{R_{\rm{d}}}/R_{\rm{d}}$ \\
     & (mag)          & ($^{\circ}$) & (mag/$''^{2}$) & (kpc)     &     & (km/s)        & (km/s)           & (kpc)             &
(km/s)                           &     & (km/s)        & (km/s)            & (km/s/kpc) \\
(1)        & (2)          &(3)& (4)           & (5)      & (6)  & (7)  &(8)& (9)      & (10)     & (11)& (12)     &(13)& (14)      \\ 
\hline
VCC 178    & -16.6$\pm$0.2 & 60$\pm$3 & 20.9$\pm$0.1 & 0.52 & a & 14$\pm$3 & 30$\pm$3 & 1.1 & 46$\pm$5 & b  & 48$\pm$5 & 73$\pm$7 & 92$\pm$12 \\
VCC 437    & -18.1$\pm$0.2 & 54$\pm$3 & 21.7$\pm$0.1 & 1.60 & a & 49$\pm$5 & 49$\pm$5 & 1.6 & 50$\pm$5 & b  & 70$\pm$5 & 70$\pm$5 & 44$\pm$5 \\
VCC 543    & -17.9$\pm$0.2 & 62$\pm$3 & 21.6$\pm$0.1 & 1.23 & a & 35$\pm$4 & 46$\pm$4 & 1.6 & 44$\pm$5 & b  & 56$\pm$5 & 68$\pm$5 & 45$\pm$5 \\
VCC 990    & -17.6$\pm$0.2 & 50$\pm$3 & 20.0$\pm$0.1 & 0.53 & a & 18$\pm$3 & 43$\pm$3 & 1.3 & 43$\pm$5 & b  & 46$\pm$5 & 80$\pm$7 & 87$\pm$11 \\
VCC 1036   & -18.3$\pm$0.2 & 69$\pm$3 & 21.4$\pm$0.2 & 1.32 & a & 32$\pm$2 & 52$\pm$2 & 2.1 & 37$\pm$4 & b  & 49$\pm$3 & 70$\pm$4 & 37$\pm$4 \\
VCC 1122   & -17.3$\pm$0.2 & 70$\pm$3 & 21.5$\pm$0.2 & 0.90 & a & 17$\pm$3 & 28$\pm$3 & 1.4 & 40$\pm$4 & b  & 43$\pm$4 & 57$\pm$5 & 48$\pm$6 \\
VCC 2019   & -17.6$\pm$0.2 & 47$\pm$3 & 20.6$\pm$0.1 & 0.73 & a & 28$\pm$3 & 42$\pm$3 & 1.1 & 41$\pm$4 & b  & 50$\pm$4 & 65$\pm$4 & 68$\pm$8 \\
VCC 2050   & -16.8$\pm$0.2 & 66$\pm$3 & 21.3$\pm$0.2 & 0.62 & a & 11$\pm$3 & 20$\pm$3 & 1.0 & 37$\pm$6 & b  & 39$\pm$6 & 51$\pm$7 & 60$\pm$10 \\
\hline
\hline
\end{tabular}
}
\tablefoot{We assumed a distance of 16.1$\pm$1.2 Mpc for all galaxies.}
\tablebib{(a)~\citet{vanZee2004a}; (b)~\citet{vanZee2004b}. }
\label{tab:Sphs}
\end{table*}

\end{appendix}

\end{document}